\documentclass[aps,prb,longbibliography,floatfix,superscriptaddress,twocolumn]{revtex4-2}
\usepackage[utf8]{inputenc}
\usepackage[canadian]{babel}
\usepackage{graphicx}
\usepackage{amsmath}
\usepackage{amsthm}
\usepackage{amsfonts}
\usepackage{amssymb}
\usepackage{mathrsfs}
\usepackage{mathtools}
\usepackage{bm}
\usepackage{textcomp,xcolor}
\usepackage[colorlinks=true,urlcolor=blue,citecolor=blue,linkcolor=blue,bookmarks=false, pdfstartview={FitH}]{hyperref}
\usepackage[caption=false]{subfig}
\usepackage[capitalize]{cleveref} 
\usepackage[normalem]{ulem} 

\hypersetup{
  hypertexnames=false,
  colorlinks=true,
  linkcolor=blue,
  anchorcolor=blue,
  citecolor=blue,
  filecolor=blue,
  urlcolor=blue,
  pdfauthor={Igor Benek-Lins; Dean Fountas; Jonathan Discenza; Saurabh Maiti},
}

\newcommand{\phantomsubfloat}[1]{
  {%
    \captionsetup[subfigure]{labelformat=empty}
    \subfloat[][]{#1}
  }%
}

\newcommand{\beq}{\begin{equation}}
  \newcommand{\eeq}{\end{equation}}
\newcommand{\bea}{\begin{eqnarray}}
  \long\def\bal#1\eal{\begin{align}#1\end{align}}
  \newcommand{\eea}{\end{eqnarray}}
\newcommand{\bse}{\begin{subequations}}
  \newcommand{\ese}{\end{subequations}}
\newcommand{\nn}{\nonumber}
\newcommand{\bwt}{\begin{widetext}}
  \newcommand{\ewt}{\end{widetext}}

\newcommand{\ve}{\varepsilon}
\newcommand{\e}{\epsilon}

\newcommand{\bk}{{\bf k}}

\let\originalleft\left
  \let\originalright\right
\renewcommand{\left}{\mathopen{}\mathclose\bgroup\originalleft}
  \renewcommand{\right}{\aftergroup\egroup\originalright}
\usepackage[ISO]{diffcoeff}[=v4]
\DeclareDocumentCommand\differential{ o g d() }{%
  \IfNoValueTF{#2}{
    \IfNoValueTF{#3}
    {\dl\IfNoValueTF{#1}{}{^{#1}}}
    {\mathinner{\dl\IfNoValueTF{#1}{}{^{#1}}\argopen(#3\argclose)}}
  }
  {\mathinner{\dl\IfNoValueTF{#1}{}{^{#1}}#2} \IfNoValueTF{#3}{}{(#3)}}
}
\newcommand{\ii}{\ensuremath{{\mkern1mu\mathrm{i}\mkern1mu}}}

\DeclareDocumentCommand\dd{}{\differential}

\DeclareMathOperator{\sgn}{sgn}

\DeclareMathOperator{\IM}{Im}
\usepackage[version=3]{mhchem}
\newcommand{\group}[1]{\ce{#1}}
\newcommand{\Bog}{\group{B_{1g}}}
\newcommand{\Btg}{\group{B_{2g}}}
\newcommand{\Aog}{\group{A_{1g}}}
\newcommand{\squarelattice}{\text{\tiny\(\square\)}}

\begin{document}

\title{Universal nonanalytic features in response functions of anisotropic superconductors}
\author{Igor Benek-Lins}
\affiliation{Department of Physics, Concordia University, Montréal, QC H4B 1R6, Canada}
\author{Dean Fountas}
\affiliation{Department of Physics, Concordia University, Montréal, QC H4B 1R6, Canada}
\author{Jonathan Discenza}
\affiliation{Department of Physics, Concordia University, Montréal, QC H4B 1R6, Canada}
\author{Saurabh Maiti}
\affiliation{Department of Physics, Concordia University, Montréal, QC H4B 1R6, Canada}
\affiliation{Centre for Research in Multiscale Modelling, Concordia University, Montréal, QC H4B 1R6, Canada}
\date{\today}

\begin{abstract}
  Nonanalytic features are interesting in physics as they carry valuable information about the physical properties of the system. These properties manifest themselves in observables containing a one- or two-particle spectral function. In this work, we use a stationary-point analysis to deduce the nonanalytic features of spectral functions that appear while computing dynamical correlation functions. We focus on the correlation functions relevant to inelastic light scattering from anisotropic superconductors and show that nodal regions of the order parameters are, quite generally, associated with linear-in-frequency scaling at low frequencies, the minima points of the order parameters are associated with step jumps, while the maxima points are associated with $\ln$ singularities. Despite this general association, we show that depending on the anisotropy of the light-scattering vertex, these features can manifest themselves as various power laws, and even not remain singular at all. We demonstrate the conditions under which these happen. We are also able to demonstrate that the association of these nonanalytic features with the extrema and nodal points of the order parameter is, in fact, derived from a more universal behaviour of functions near parabolic-like and saddle-like stationary points. We provide a general prescription that maps the universal behaviour of systems near such stationary points to different scenarios. The approach is readily extendable to other types of spectral functions and we exemplify it by also analyzing the density of states on a square lattice.
\end{abstract}

\maketitle

\tableofcontents

\section{Introduction}\label{Sec:Introduction}
An important quantity in the study of solid-state materials is the spectral function (or functions derivable from it) which tracks the possible energies and momenta at which the system can be excited. The excitation itself can be single-particle-like (fermionic) or multi-particle-like (bosonic and collective). Probes like angle-resolved photoemission spectroscopy~\cite{Damascelli2003}, scanning tunnelling spectroscopy~\cite{Yazdani2016} provide access to the former, while probes like infrared absorption~\cite{Basov2011} and inelastic scattering light scattering (via Raman scattering~\cite{devereaux_hackl:2007:InelasticLightScatteringCorrelated} or resonant inelastic X-ray scattering~\cite{Ament2011}) provide access to the latter. Although, improvements in the experimental techniques have allowed the so-called single-particle probes to also access many-particle collective phenomena~\cite{Mori2023,Lee2023,Boschini2024}.

Studying the spectral function helps us deduce important quantum properties of the material~\cite{coleman:2015:IntroductionManyBodyPhysics,mahan:2000:ManyParticlePhysics}. However, being able to accurately interpret various features in the data has proved challenging. Some examples of controversial interpretations of data from superconductors~\footnote{We mention here that the interpretations were almost always presented in accordance with the best data and theory available at that time. This point is being raised not to criticize the works but to highlight the need to understand the phase space of responses in different scenarios} are (i) the association of THz oscillations to the Higgs mode~\cite{Matsunaga2014} (when the response is expected to be dominated by pair-breaking excitations~\cite{Cea2016THG}); (ii) the non-suppressed $\Aog$ response in cuprate superconductors despite theory arguing for a heavily screened response~\cite{Martinho2004, devereaux_hackl:2007:InelasticLightScatteringCorrelated}; (iii) the confusion in association of spectral features in electronic Raman scattering to pair-breaking features~\cite{bohm_etal:2014:BalancingActEvidenceStrong,wu_etal:2017:SuperconductivityElectronicFluctuationsCe} and Bardasis--Schrieffer collective modes~\cite{kretzschmar_etal:2013:RamanScatteringDetectionNearlyDegenerate,bohm_etal:2018:MicroscopicOriginCooperPairing}. One major reason for ambiguities in the interpretation of experimental data is the lack of insight from theory, which in turn is due to the difficulty in performing analytical calculations. While one can certainly invoke numerical methods, these calculations are expensive and one seldom puts in the resources to calculate many-body corrected responses after getting an estimate for the ground state of the system (which is already computationally expensive). There are notable exceptions~\cite{Klein1984, Devereaux1995,Scalapino2009,Khodas2014,Cea2016,Maiti2016}, but they still assume a simple toy model to produce a workable formula that can at least qualitatively explain the new phenomena seen in the experiments. These simple models, by design, leave out important factors such as the details of the anisotropy of the electronic state and/or the established order parameter. A general formulation was proposed in Ref. \cite{Maiti2017} to counter this. Nevertheless, the specific role of anisotropy was not explicitly explored. It is not clear, \emph{a priori}, if it can introduce fundamentally new/unique signatures. If they do, it is not clear if these features could be unified under one or classified into different origins. Also, it is not clear if a feature found numerically is specific to the scenario it was calculated under or if some general property controls its manifestation. This is certainly not a new concern. One of the early studies exploring the experimental consequence of anisotropy arising from a mixed symmetry order parameter state was explored in Ref. \cite{Devereaux1995}, where the authors found interesting features such as a step jump, $\ln$ peaks and different power law onsets. However, it was not explained if these features were interconnected or if there was some universality to this.

Given the recent rise in the study of anisotropic superconductivity in popular 2D quantum materials such as orbital-selective correlated \ce{FeSe}~\cite{Sprau2017}, rhombohedral trilayer graphene~\cite{Zhou2021,Chatterjee2022,Levitan2024}, and possibly even in the ferromagnetic \ce{U}-based superconductors~\cite{Aoki2019}, developing an intuition for how anisotropic systems could respond will prove to be immensely helpful. In this work, we do not aim to resolve any material-specific debates, but aim to provide an answer to the general question of the role of anisotropies and their manifestations in response functions. This article serves two purposes: (i) it reminds the reader of an old trick of analyzing stationary points and shows how to use it to deduce spectral features with minimal calculations, and (ii) to classify various spectral features, discussing their origins, their universality and even the different manifestations of the same feature in different probes. The first result identifies the relevance of two types of stationary points in the pole: the parabolic $x^2+y^2$-like (where $x$ and $y$ are the integration variables) and the saddle $x^2-y^2$-like. It then outlines a prescription for performing a local asymptotic analysis that helps extract the unique features of a response. The second result applies this prescription to the case of calculating bare Raman correlation functions in superconductors that links a given type of the stationary point to nonanalytic features in the response such as: a step jump $\sim \Theta (\Omega-\Omega_0)$, an onset with a power law $(\Omega-\Omega_0)^n$, with $n\ge0$, and a $\ln$ singularity $\sim\ln|\Omega-\Omega_0|$. Here, $\Theta$ is the Heaviside step function, $\Omega$ is a spectral frequency and $\Omega_0$ is the location of the nonanalyticity. The exponent $n$ is shown to be directly related to the number of zeros at the pole of the integration. Quite generally, it is shown that parabolic-stationary points lead to step jumps, which can be modified to onset-like behaviour when dressed by zeros in an appropriate coupling vertex of a probe. The saddle-stationary points lead to a $\ln$ nonanalyticity, which can also be modified through the coupling vertex of a probe. One can also universally associate these nonanalytic behaviours with the minimum, maximum or nodal regions of the order parameter of an anisotropic superconductor. A single correlation function can have multiple locations at which it is nonanalytic. Our result not only identifies all of them, but allows one to combine them to get the behaviour of the full response function.

The rest of the text is organized as follows. In Sec.~\ref{Sec:Prescription} we state the prescription and describe the rationale for it. It also lists results of some common integrations from which the various features we find originate. In Sec.~\ref{Sec:Applications2} we apply the prescription to the case of Raman scattering from superconductors for various structures of the order parameter. We extract the universal features and discuss their manifestations for different scattering geometries of the experiment. In Sec.~\ref{Sec:Applications1}, as another example of the applicability of the prescription, we discuss the density of states of electrons in free space and in a lattice in various spatial dimensions. In Sec.~\ref{Conclusion} we summarize our results and present other scenarios where the results and the prescription could be used. Finally, in the appendices we present a derivation of the results of integration over poles that give the universal features discussed in this article.

\section{Prescription for asymptotic analysis}\label{Sec:Prescription}
Singular features in an integral only arise from nonanalytic/singular behaviour of the integrand, examples of which will be demonstrated later in the article. The simplest form of singularity of the integrand would be due to a simple pole of the form $x-z$ in the denominator. Here, $x$ is the integration variable and $z$ is some external parameter. The integrations of interest to us (and generally in physics) will also include a regulator parameter $\eta\rightarrow 0$ such that the pole takes the form $x-z-\ii\eta$. This prevents a true singularity and also helps identify the relevant branches near the singularities in the complex plane. This can be generalized from simple poles to arbitrary forms of the poles, $f(x)-z-\ii\eta$. And this can further be extended to two dimensions with a general form $f(x,y)-z-\ii\eta$.

One can then exploit the fact that the fast divergence of the pole can be isolated from the rest of the integrand by casting the latter into the form $R(x_0,y_0)/[f(x,y)-z-\ii\eta]$, where $R(x_0,y_0)$ is the ``residue'' of the integrand at the pole. We then integrate across the singularity (that is regulated by $\eta$). However, in more than one dimension, this integration is not over a point, but over either a line or a surface. Thus, it would seem that $(x_0,y_0)$ should correspond to a contour of points. However, if we are interested in universal features of the integral introduced by the singularities of an integrand, then such features only arise from specific regions of the contour of poles in the integration space. These regions are identified by the \textit{stationary points} of the denominator, here $f(x,y)$, which are found by solving $\partial_xf(x_0,y_0)=0$ and $\partial_yf(x_0,y_0)=0$. Around these stationary points, the denominator takes the form
$f(x_0+x,y_0+y)=f_0+x^2\partial^2_xf_0/2+y^2\partial^2_yf_0/2+xy\partial_x\partial_yf_0$,
where $f_0\equiv f(x_0,y_0)$. The universal features of the integral can be deduced by just locally integrating across the pole around the stationary points of the integrand's denominator. These universal features completely dictate the asymptotic behaviour of the integral local to a singular point whose location is a function of the external parameter $z$. Such approximation techniques have been well explored in quantum mechanics in the form of the WKB approximation~\cite{Griffiths:2018:IntroductionQM} and in quantum field theory in the form of the saddle-point approximation~\cite{FradkinBook2021}. In such cases, this approximation is invoked to offer a rule to make progress towards understanding the behaviour of otherwise open/intractable problems. We direct the reader to Ref.~\cite{FlajoletBook2009} for a more comprehensive consideration of saddle-point approximations. In this work, we revert back to the idea of utilizing stationary points, but this time to deduce interesting and universal features of response functions that are ubiquitous in condensed-matter physics. These often get computed in a brute force manner that prevents one from anticipating universality in the nature of the responses. In fact, as we shall demonstrate, almost the entire response can be presented with reasonable accuracy by just collecting these universal features at various singular points. Besides identifying the universal features, we can also identify the region of the integration space that is responsible for these features. How this information helps us will also be demonstrated in this article.

In precise terms, the proposed prescription calls for the following steps:
\begin{itemize}
  \item Identify the stationary points $(x_0,y_0)$, and expand locally around the stationary point.
  \item  If the integrand has a singularity [of the form $f(x,y)-z$], choose a local integration region that covers the pole and integrate over the it.
  \item If the integral is singular, it will be at the value of $z=f_0$, the value of $f$ at the stationary point $(x_0,y_0)$.The result depends on $z-f_0$ and establishes the local $z$-dependence of the integration. The integration itself can be done by separating the fast varying singularity from the slow varying ``residue'', which is evaluated at the pole. One can further identify interesting regimes as a function of the external parameter $z$ by considering cases when the pole is near zero, i.e., $z\rightarrow f_0$, or when $z\gg f_0$. The region of the local integration will move to include the pole.
  \item If there are more stationary points, repeat the same exercise and add all the contributions to get the final $z$-dependent result.
\end{itemize}
In Secs.~\ref{Sec:Applications2} and \ref{Sec:Applications1} we will demonstrate these ideas through several examples.

Before we proceed, some general statements are in order. When we refer to a universal feature, we mean that (i) the result remains finite even in the regulator limit $\eta\rightarrow 0$, and (ii) the result is independent of the integration limits. Certainly, there are also non-universal contributions to the integral. However, these will hardly be necessary to capture the characteristic $z$-dependence of the integral. This is particularly relevant for the cases where we calculate response functions. In these cases, one is usually interested in the imaginary parts of the integrations over poles shifted to the complex plane by the imaginary regulators (see, for example, the fluctuation--dissipation theorem~\cite{coleman:2015:IntroductionManyBodyPhysics}). The response, being proportional to the imaginary part, only arises from around the poles due to the imaginary regulator \(\eta\). In the other regions of integration, the imaginary regulator can be ignored, leading to contributions to the integral that are real. Thus, we do not have to worry about the entire region of integration. Another aspect that can be exploited is that the expansion term with $xy$ can also be removed by an appropriate change of variable in linear combinations of $x$ and $y$. This change of variable usually affects the limits of the integration, but since the universal features, when they exist, do not depend on them, it allows us to only consider expansion with terms $x^2$ and $y^2$. We summarize the general forms of such integrations next.

\subsection{List of common integrations around poles}
In 2D, arbitrary coefficients before $x^2$ and $y^2$ can always be scaled and cast into the combinations $x^2+y^2$ and $x^2-y^2$. To this effect, we only need the following integrals:
\begin{align}
  \IM\left(\int \dd{x}\frac{1}{x-z-\ii\eta}\right)&=\pi \sgn(\eta);\label{eq:1D1}\\
  \IM\left(\int \dd{x}\frac{1}{x^2-z-\ii\eta}\right)&=\frac{\pi}{2\sqrt{z}}\Theta(z) \sgn(\eta);\label{eq:1D2}\\
  \IM\left(\int \dd{x}\dd{y}\frac{1}{x^2+y^2-z-\ii\eta}\right)&=\pi^2 \Theta(z) \sgn(\eta);\label{eq:2D1}\\
  \IM\left(\int \dd{x}\dd{y}\frac{1}{x^2-y^2-z-\ii\eta}\right)&=\pi\ln\left(\frac{\lambda}{|z|}\right) \sgn(\eta).\label{eq:2D2}
\end{align}
Here, $\sgn$ is the sign function, and the parameter $\lambda$ is a non-universal constant that is immaterial near the singularity $z\rightarrow 0$, but is necessary to make the argument of $\ln$ dimensionless. Their derivations are discussed in detail in Appendices~\ref{Sec:App1}--\ref{Sec:App4}. In writing $z$ above we have subsumed the terms of the type $z-f_0$ and its appropriate scalings. Note that Eq. (\ref{eq:1D1}) is not an integration associated with a stationary point, but we included it as it is a common singularity one encounters. But observe that the result is not $z$-dependent consistent with the fact that non-stationary points do not introduce $z$-dependent singular features, vide Appendix~\ref{Sec:App5}.

In 1D cases [\cref{eq:1D1,eq:1D2}], the integrals are mostly dictated  by the integration around the simple poles at points $z$ and $\pm\sqrt z$. In 2D [\cref{eq:2D1,eq:2D2}], the simple poles get extended to lines. As justified above, it would suffice to focus on the extension to a circle ($x^2+y^2=z$) and a hyperbola ($x^2-y^2=z$). As far as the labelling is concerned, instead of referring to the contours of the pole, one may also refer to the geometry of the manifold of the integration space and call them \emph{parabolic}-stationary and \emph{saddle}-stationary points, respectively. We remind the reader that $x$ and $y$ should be mapped back to the original space to infer the region of integration in the original variables of the problem.

From these basic integrations, we can easily deduce the result for more involved integrands. For example, if the numerator contained a weight factor $w(x,y)$ that had zeros within the radius of integration, it would affect the manifestation of the singularity after the integration in ways that can even effectively remove a previously universal contribution. See corollaries of Appendices~\ref{Sec:App3} and \ref{Sec:App4} where this is mathematically demonstrated. Examples of such cases will also be discussed in Sec.~\ref{Subsec:Selection}. Although we have restricted ourselves to 1D and 2D integrations in this work, one can certainly extend this to 3D, but we shall not cover these cases in this article.

\section{Universal features in dynamical correlation functions}\label{Sec:Applications2}
The response from a causal system is modeled as a time-correlation function where one has to extend the frequency variable $\Omega$ infinitesimally to the complex plane ($\Omega\rightarrow\Omega+\ii\eta,~\eta\rightarrow0^+$). The physical response is then proportional to the imaginary part of such a correlation. For the purpose of this work, we will pick the response functions of superconductors (SCs). It will provide us with many variations that are experimentally relevant in current research on 2D and quasi-2D materials.  To this effect, let us consider the correlation function associated with electronic Raman spectroscopy~\cite{Dierker1983,Devereaux1995,Chubukov1999}
\bwt
\begin{equation}\label{eq:chi1}
  \chi_C(\Omega)\stackrel{\hphantom{\!\!{}^{+}}\eta\rightarrow0^+}{=}\int_0^{2\pi}\frac{\dd{\theta}}{2\pi}\gamma^2_C(\theta)\int_0^{\Lambda_{\text{b}} }\dd{\ve}\frac{1}{\sqrt{\ve^2+\Delta_\theta^2}}\frac{\Delta_\theta^2}{\ve^2+\Delta_\theta^2-(\Omega+\ii\eta)^2/4}.
\end{equation}
\ewt
Here, $\Delta_\theta$ is the order parameter of the superconductor with $\theta$ being the angle along the Fermi surface. The label $C\in\{\Aog, \Bog, \Btg\}$ represents the irreducible representations (irreps) of the square lattice (this choice is not significant, but we pick it for definiteness and relevance towards experiments \cite{devereaux_hackl:2007:InelasticLightScatteringCorrelated,lazarevic_hackl:2020:FluctuationsPairingFebasedSuperconductors}). The form factors of \(\gamma_{C}\) (Raman vertex) in these irreps are $\gamma_{\Aog}=1$, $\gamma_{\Bog}=\cos(2\theta)$ and $\gamma_{\Btg}=\sin(2\theta)$. Furthermore, $\Omega$ is the energy transferred to the system and the spectral frequency at which the measurement took place, $\Lambda_{\text{b}}$ is the bandwidth of the band, and the integration variable $\varepsilon$ represents the energy measured relative to the Fermi energy. This bandwidth is typically larger than any low-energy scale (here $\Delta_\theta$). Lastly, the relevant quantity to model a measurement is $\IM[\chi_C(\Omega)]$. We alert the reader that calculating $\chi_C(\Omega)$ (known as the \emph{bare correlation function}) is merely the first step towards calculating the full response which includes many-body effects~\cite{Klein1984,Devereaux1995,devereaux_hackl:2007:InelasticLightScatteringCorrelated,Marciani2013,Cea2016,Maiti2017,Benek-Lins2024}. Our goal here is not to model the true response, but to demonstrate how the main features of a response function could be extracted without explicit computations. In what follows next, we will start with $C=\Aog$ and investigate the bare Raman correlation function~\footnote{We inform the reader here that when many-body effects are accounted for, this response is actually null~\cite{Cea2016}, but this is not of subject of concern for this article.} for superconductors with distinct order parameter structures. We will subsequently switch to the other irreps.

\subsection{Raman correlation function in an isotropic $s$-wave superconductor}
Here, the order parameter is angle independent: $\Delta_\theta=\Delta_{0} = \text{constant}$. In this text, any constant parameters we introduce (e.g., $\Delta_0$ here), will be chosen to be positive for definiteness. For $C=\Aog$, the response is then obtained from the following integration
\begin{align}\label{eq:chis}
  \chi_{\Aog}^{\text{$s$-wave}}(\Omega)\stackrel{\hphantom{\!\!{}^{+}}\eta\rightarrow 0^+}{=}\int_0^{\Lambda_{\text{b}} }\dd{\ve} &\frac{1}{\sqrt{\ve^2+\Delta_0^2}}\nonumber\\
                                                                                                                                  & \times \frac{\Delta_0^2}{\ve^2+\Delta_0^2-(\Omega+\ii\eta)^2/4}.
\end{align}
In this effective 1D integration, the stationary point is at $\varepsilon=0$ and the singularity falls under the category of \cref{eq:1D2} with $z=(\Omega/2)^2-\Delta_{0}^2$. The integral is weighted with  $w(\ve)=\Delta_{0}^2/\sqrt{\ve^2+\Delta_{0}^2}$ which does not diverge and will thus provide a ``residue'' that will be evaluated at the pole. Since the pole is at $\ve=\sqrt{(\Omega/2)^2-\Delta_{0}^2}$, using \cref{eq:1D2} we can readily write down the final result as
\begin{align}\label{eq:chis2}
  \IM[\chi_{\Aog}^{\text{$s$-wave}}(\Omega)]
  =&\frac\pi2\frac{1}{\Omega/(2\Delta_0)}\frac{1}{\sqrt{[\Omega/(2\Delta_0)]^2-1}}\nonumber\\
   &\times\Theta(\Omega-2\Delta_0).
\end{align}
To compare this with the actual result, we first break down this result to the following cases for the local asymptotic forms:
\begin{align}\label{eq:results}
  &\IM[\chi^{s\text{-wave}}_{\Aog}(\Omega)]\nonumber\\
  &=\begin{cases}
    \displaystyle 0 &\text{for }\Omega\ll2\Delta_{0},\\
    \displaystyle \frac{\pi}{2}\frac{1}{\sqrt{[\Omega/(2\Delta_{0})]^2-1}}\Theta(\Omega-2\Delta_{0}) &\text{for }\Omega\approx2\Delta_{0},\\[2ex]
    \displaystyle \frac{\pi}{2[ \Omega/(2\Delta_{0}) ]^2} &\text{for }\Omega\gg2\Delta_{0}.
  \end{cases}
\end{align}
The exact result of integrating \cref{eq:chis} with $\Lambda_{\text{b}}\rightarrow\infty$ is
\begin{equation}
  \label{eq:exact_result}
  \chi(\Omega)=\frac{\arcsin[\Omega/(2\Delta_{0})]}{\Omega/(2\Delta_{0})}\frac{1}{\sqrt{1-[\Omega/(2\Delta_{0})]^2}}.
\end{equation}
\begin{figure}
  \centering
  \includegraphics[width=\linewidth]{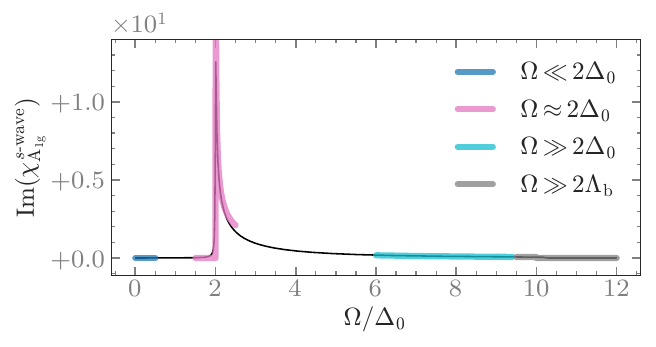}
  \caption{\textbf{Reproducing asymptotic features of a constant gap}: Comparison of asymptotic features obtained from our prescription (light thick lines), against the exact result (black thin line) for an isotropic $s$-wave superconductor for the irrep $C=\Aog$. The $1/\sqrt{\Omega/(2\Delta_0)-1}$ feature and the $1/\Omega^2$ fall off are accurately reproduced. The drop to zero at $\Omega\sim2\Lambda_{\text{b}}$ is exact for \(\Lambda_{\text{b}}/\Delta_{0} \to \infty\), but here it is offset due to its finiteness, \(\Lambda_{\text{b}} = 5\Delta_0\).}
  \label{fig:s_isotropic}
\end{figure}
The plot of the exact result and the asymptotic forms in \cref{fig:s_isotropic} shows how well our prescription reproduces the exact result. Observe that the entire result comes from the neighbourhood of the pole at $\varepsilon\sim \sqrt{(\Omega/2)^2-\Delta_0^2}$, irrespective of $\Omega$. As stated in the prescription, the singularity of the final result is determined by setting the pole to zero. Here, this happens at $\Omega=2\Delta_0$, which is corroborated in \cref{fig:s_isotropic}.

In fact, we can also address the step drop at $\Omega\sim 2\Lambda_{\rm b}$ within our prescription. To do so, introduce a dimensionless parameter $x$ such that $\Omega=2\Lambda_{\text{b}}(1-x)$, $x>0$. Note that the pole contribution here comes from $\ve\sim\Omega/2\gg\Delta_{0}$. The integration then can be expressed as
\begin{align}\label{eq:Reg3a}
  &\IM[\chi^{s\text{-wave}}_{\Aog}(\Omega)]\nonumber\\
  &=\IM\left[\int_{\Omega/2-\delta}^{{\rm min}( \Omega/2+\delta, \Lambda_{\text{b}} ) }\dd{\ve}\frac{1}{\Omega/2}\frac{\Delta_{0}^2}{\ve^2-(\Omega+\ii\eta)^2/4}\right]\nn\\
  &\approx\frac{\Delta_{0}^2}{2\Lambda^2_{\rm b}}\IM\left(\int_{-\delta}^{{\rm min}( \delta, \Lambda_{\text{b}} x ) }\dd{\ve}\frac{1}{\ve-\ii\eta}\right)\nn\\
  &=\frac{\pi}{2\Lambda_{\text{b}}^2/\Delta_{0}^2}\Theta(\Lambda_{\text{b}}x),~\text{from \cref{eq:1D1}}\nn\\
  &=\frac{\pi}{2\Lambda_{\text{b}}^2/\Delta_{0}^2}\Theta(2\Lambda_{\text{b}}-\Omega).
\end{align}
In the penultimate line, the $\Theta$ function ensures that the upper limit is positive guaranteeing the inclusion of the pole within the integration range. This result simply reestablishes the physical fact that there can be no response from states beyond the bandwidth. This cut-off of the response at the scale of the bandwidth is independent of the low-energy structure of $\Delta_\theta$ and will apply to all cases below. Thus, we will omit this detail from the remaining discussions.

\subsection{Raman correlation function in an anisotropic $d$-wave superconductor}\label{Subsec:dwave}
This case corresponds to $C=\Aog$ and $\Delta_\theta=\Delta_2\cos(2\theta)$ in \cref{eq:chi1}. Such an order parameter is typical for cuprates~\cite{CHUBUKOV2024}. We can change the integration limits on $\theta$ from $[0,2\pi)$ to the first quadrant, \([0, \pi/2]\), and multiply the integral by a factor of 4. The stationary point on the $\varepsilon$-axis is $\varepsilon=0$ and on the $\theta$-axis is found from $\partial_\theta\Delta_\theta^2=0$ which gives $\theta = 0,\pi/4,\pi/2$ in the first quadrant. At the points $\theta=0,\pi/2$, $\Delta^2_\theta$ has a maximum, while at $\theta=\pi/4$, $\Delta^2_\theta$ has a minimum. Knowing that the imaginary part would only come from integration around the poles, the limits in \cref{eq:chi1} can be restricted as follows:
\bwt
\begin{align}\label{eq:chi2}
  \IM[\chi^{\text{$d$-wave}}_{\Aog}(\Omega)]&=\frac2\pi \IM\left[\int_0^{\pi/2}\dd{\theta}\int_0^{\Lambda_{\text{b}} }\dd{\ve}\frac{1}{\sqrt{\ve^2+\Delta_\theta^2}}\frac{\Delta_\theta^2}{\ve^2+\Delta_\theta^2-(\Omega+\ii\eta)^2/4}\right]\nn\\
                                            &=\frac2\pi \IM\left[\left(\int_0^{\delta}+\int_{\pi/4-\delta}^{\pi/4+\delta}+\int^{\pi/2}_{\pi/2-\delta}\right)\dd{\theta} \int_0^{\Lambda_{\text{b}} }\dd{\ve}\frac{1}{\sqrt{\ve^2+\Delta_\theta^2}}\frac{\Delta_\theta^2}{\ve^2+\Delta_\theta^2-(\Omega+\ii\eta)^2/4}\right].
\end{align}
\ewt
In the second line, we have introduced an arbitrary partition at $\delta$ to separate the stationary-point regions. The $\delta$'s can be different in different regions, but since we already established that universal features should not depend on the limits, we set them equal for brevity. In fact, with appropriate change of variables ($\theta\rightarrow\theta-\pi/2$), the third interval can be combined with the first one, which allows us to write
\bwt
\begin{equation}\label{eq:chi3}
  \IM[\chi^{\text{$d$-wave}}_{\Aog}(\Omega)]=\frac2\pi \IM\Bigg[\Big(\underbrace{\int_{-\delta}^{\delta}}_{(a)}+\underbrace{\int_{\pi/4-\delta}^{\pi/4+\delta}}_{(b)}\Big) \dd{\theta} \int_0^{\Lambda_{\text{b}} }\dd{\ve}\frac{1}{\sqrt{\ve^2+\Delta_\theta^2}}\frac{\Delta_\theta^2}{\ve^2+\Delta_\theta^2-(\Omega+\ii\eta)^2/4}\Bigg].
\end{equation}
\ewt
Let us first start with the term $(a)$, which corresponds to the maximum point. The expansion around the stationary point yields $\Delta_\theta\approx\Delta_2-2\Delta_2\theta^2$ which allows us to write
\begin{align}\label{eq:chi3newa}
  (a)=\IM\Bigg[&\int_{-\delta}^{\delta}\dd{\theta} \int_0^{\Lambda_{\text{b}} }\dd{\ve} \frac{1}{\sqrt{\ve^2+\Delta_2^2}}\nonumber\\
               & \times\frac{\Delta_2^2}{\ve^2+\Delta_2^2-4\Delta_2^2\theta^2-(\Omega+\ii\eta)^2/4}\Bigg]
\end{align}
This integration takes the form of \cref{eq:2D2}. The pole term is $\sqrt{(\Omega/2)^2-\Delta_2^2}$ and a singular behaviour may arise when this term $\rightarrow0$, i.e., when $\Omega\rightarrow 2\Delta_2$. When $\Omega$ is around this region, the pole and hence the integration variables are forced to be small. This allows the integration to be carried out as
\begin{align}\label{eq:aterm}
  (a)&= \Delta_2 \IM\Bigg[\int_{-\delta}^{+\delta} \dd{\theta}\int_0^{\Lambda_{\rm b}}\dd{\e} \nonumber \\
     & \hphantom{= \Delta_2 \IM\Bigg[}\times\frac{1}{\e^2+\Delta_2^2(1-4\theta^2)-(\Omega+\ii\eta)^2/4}\Bigg]\nn\\
     &=\frac{1}{4} \IM\left\{\int \dd{x}\dd{y}\frac{1}{x^2-y^2-2\Delta_2^2[\Omega/(2\Delta_2)-1]-\ii\eta}\right\}\nn \\
     &=\frac{\pi}{4}\ln\left[\frac{\lambda}{|\Omega/(2\Delta_2)-1|}\right].
\end{align}
As in \cref{eq:2D2}, due to the logarithm, a scale $\lambda$ gets introduced in the problem that is formally not determined in this method. But this parameter is not significant near the singular behaviour.

For the term \((b)\), which corresponds to the minimum, we have $\Delta_\theta\approx2\Delta_2\varphi$, where $\varphi\equiv\theta-\pi/4$. The integration can then be cast as
\begin{align}\label{eq:chi3newb}
  (b)=\IM\Bigg[&\int_{-\delta}^{\delta}\dd{\varphi} \int_0^{\Lambda_{\text{b}} }\dd{\ve} \frac{1}{\sqrt{\ve^2+4\Delta_2^2\varphi^2}} \nonumber \\
               &\times\frac{4\Delta_2^2\varphi^2}{\ve^2+4\Delta_2^2\varphi^2-(\Omega+\ii\eta)^2/4}\Bigg].
\end{align}
Here, the pole term is $(\Omega/2)^2$ and hence any interesting feature can only be expected near $\Omega\rightarrow 0$. This integration is of the type \cref{eq:2D1}, but with a weight function multiplying the singular integrand which has a zero. To evaluate this, we can switch to polar coordinates \((r, \xi)\) and follow the steps outlined in the corollary of Appendix~\ref{Sec:App4} to get:
\begin{align}\label{eq:bterm}
  (b)&= \frac{1}{4\Delta_2} \IM\left[\int_0^\lambda r\dd{r}\int_0^{2\pi} \dd{\xi}\frac{1}{r}\frac{r^2\cos^2(\xi)}{r^2-(\Omega+\ii\eta)^2/4}\right]\nn\\
     &= \frac{\pi}{4\Delta_2} \IM\left[\int_0^\lambda \dd{r}\frac{r^2}{r^2-(\Omega+\ii\eta)^2/4}\right]\nn\\
     &=\frac{\pi}{4\Delta_2}\frac{\pi(\Omega/2)^2}{2(\Omega/2)},~\text{from \cref{eq:1D2}}\nn\\
     &=\frac{\pi^2}{8}\left(\frac{\Omega}{2\Delta_2}\right).
\end{align}
Observe that the result is still consistent with \cref{eq:2D1}. We have used $\Omega>0$ above and hence the $\Theta$-function is not shown, and the additional factor of $\Omega$ in the result (as opposed to a constant) arose due to the weight function $\sim \varphi^2/\sqrt{\varepsilon^2+\varphi^2}$ associated with the singularity. In fact, every additional zero in terms of the integration variable will add a power of the pole to the result. Here, since the combination $\varphi/\sqrt{\varepsilon^2+\varphi^2}$ is dimensionless, the expression $\varphi^2/\sqrt{\varepsilon^2+\varphi^2}$ only has one zero.

Thus, our prescription allows us to conclude that from the $(a)$ term, which is a \textit{maximum point} and comes from $\theta\sim 0,\pi/2$, we get a feature near $\Omega-2\Delta_2\rightarrow0$ of the form $\propto\ln|\Omega-2\Delta_2|$, while from the $(b)$ term, which is a minimum point and comes from $\theta\sim\pi/4$, we get a feature near $\Omega\rightarrow0$ of the form $\propto\Omega$. Since the two regions contribute at different $\Omega$'s, their additions do not overlap.

\begin{figure*}
  \centering
  \phantomsubfloat{\label{fig:A1g_results_a}}\phantomsubfloat{\label{fig:A1g_results_b}}\phantomsubfloat{\label{fig:A1g_results_c}}\vspace{-2\baselineskip}
  \includegraphics[width=1.0\linewidth]{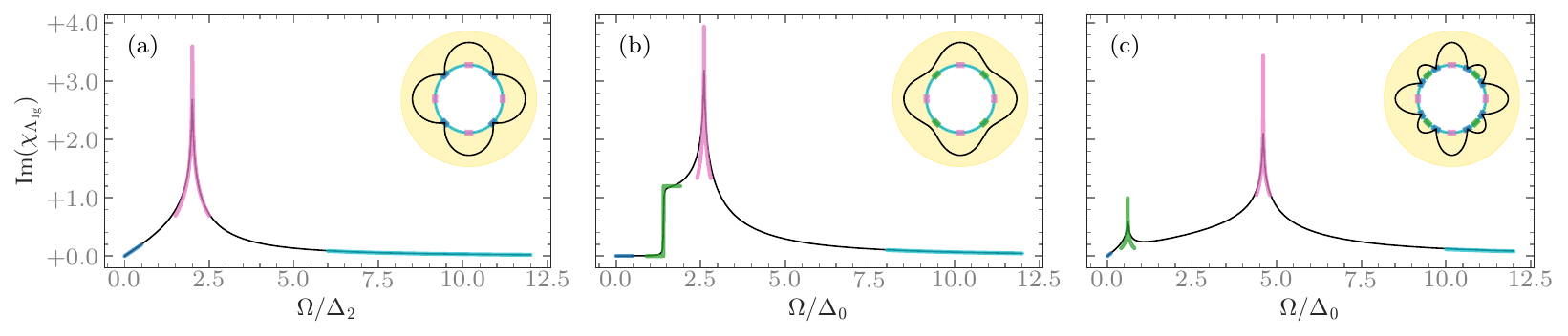}
  \caption{\textbf{Asymptotic features from anisotropic gaps}: The asymptotic features from our prescription against  exact calculations for the bare Raman correlation function for $C=\Aog$ for (a) a $d$-wave, (b) a nodeless anisotropic \(s\)-wave and (c) a nodal anisotropic \(s\)-wave SC. In the inset we show the associated order parameter (black line), the form factor of the vertex $\gamma_{\Aog}$ in shaded yellow (which is a constant here), and in various colours we represent the regions of the Fermi surface that yield the asymptotic features in the plot. Quite generally, the large frequency tail comes from all of the Fermi surface, the $\ln$ peaks come from order-parameter maxima regions, the step jumps come from the minima regions of the order parameter. The nodal regions of the order parameter contribute to the linear slope at low frequencies.}
  \label{fig:A1g_results}
\end{figure*}

Finally, in the large frequency limit where $\Omega\gg2\Delta_2$, the pole integration of \cref{eq:chi2} essentially gets reduce to a 1D form. This is because $\Delta_\theta$ is incapable of providing a large enough value to integrate around the pole (but $\ve$ can). The integration looks similar to the $s$-wave case, but for the angle-averaged prefactor:
\begin{align}\label{eq:largeW}
  & \IM[\chi^{\text{$d$-wave}}_{\Aog}(\Omega)] \nonumber\\
  & =\frac2\pi \IM\left[\int_0^{\pi/2} \dd{\theta}\int_0^{\Lambda_{\text{b}} }\dd{\ve}\frac{1}{\Omega/2}\frac{\Delta_\theta^2}{\ve^2-(\Omega+\ii\eta)^2/4}\right]\nn\\
  &=\frac{\Delta_2^2}{\Omega}\IM\left[\int \dd{\ve}\frac1{\ve^2-(\Omega+\ii\eta)^2/4}\right]\nn\\
  &=\frac{\Delta_2^2}{\Omega}\frac{\pi}{\Omega},~\text{from \cref{eq:1D2}}\nn\\
  &=\frac{\pi}{4}\frac1{[\Omega/(2\Delta_2)]^2}.
\end{align}
Collecting all these results, we find
\begin{align}\label{eq:d_aog_result}
  \IM[\chi^{d\text{-wave}}_{\Aog}(\Omega)]&=\begin{cases}
    \displaystyle \frac{\pi}{4}\left(\frac{\Omega}{2\Delta_2}\right) &\text{for }\Omega\ll2\Delta_2,\\[2ex]
    \displaystyle \frac{1}{2}\ln\left[\frac{\lambda}{|\Omega/(2\Delta_2)-1|}\right] &\text{for }\Omega\approx2\Delta_2,\\[2ex]
    \displaystyle \frac{\pi}{4}\frac1{[\Omega/(2\Delta_2)]^2} &\text{for }\Omega\gg2\Delta_2.
  \end{cases}
\end{align}
In addition to these forms, we also know exactly which parts of the Fermi surface are responsible for these contributions. These analytical results are plotted in \cref{fig:A1g_results_a} against an exact computation for the $d$-wave system. The parameter $\lambda$ is calibrated to align the $\ln$ peak to the exact curve. Besides the excellent agreement, we also see the regions of the Fermi surface in different colours contributing to the corresponding features in the same colour. Note that the high-frequency tail comes from all around the Fermi surface, the $\ln$ feature comes from the maximum of the order parameter, and the $\propto\Omega$ feature from the nodal regions. In an exact calculation (done numerically, for example), it would not be apparent what features to expect and which parts of the Fermi surface would be responsible for the features.

\subsection{Raman correlation function in an anisotropic $s$-wave superconductor}
In real systems, lattice anisotropies make the order parameter anisotropic but still consistent with the symmetries of the lattice. One such example is the extended $s$-wave state (found in \ce{Fe}-based superconductors~\cite{Maiti2011}) that has the form factor $\Delta_\theta=\Delta_0+\Delta_4\cos(4\theta)$. The extra parameter $\Delta_4$ introduces two regimes: the nodeless regime where $\Delta_4<\Delta_0$ and the order parameter remains positive, and the nodal regime where $\Delta_4>\Delta_0$ and the order parameter crosses zero.

In the nodeless case, the stationary points with respect to $\theta$ are the minimum with $\Delta_{\rm min}=\Delta_{\theta=\pi/4}=\Delta_0-\Delta_4$ and the maximum with $\Delta_{\rm max} = \Delta_{\theta=0,\pi/2}=\Delta_0+\Delta_4$. See inset of \cref{fig:A1g_results_b} for the structure of the order parameter. Like in \cref{eq:chi2,eq:chi3}, the angle integration is effectively split into two parts: an integration around $\theta=0$ and another around $\theta=\pi/4$:
\bwt
\begin{align}\label{eq:chian}
  \frac{\pi}{2}\IM[\chi_{\Aog}(\Omega)]&= \IM\left[\left(\int_{-\delta}^{+\delta}+\int_{\pi/4-\delta}^{\pi/4+\delta}\right) \dd{\theta} \int_0^{\Lambda_{\text{b}} }\dd{\ve}\frac{1}{\sqrt{\ve^2+\Delta_\theta^2}}\frac{\Delta_\theta^2}{\ve^2+\Delta_\theta^2-(\Omega+\ii\eta)^2/4}\right]\nn\\
                                       &=\underbrace{\IM\left[\int_{-\delta}^{+\delta} \dd{\theta} \int_0^{\Lambda_{\text{b}} }\dd{\ve}\frac{1}{\sqrt{\ve^2+\Delta_{\rm max}^2}}\frac{\Delta_{\rm max}^2}{\ve^2+\Delta_{\rm max}^2-16\Delta_4\Delta_{\rm max}\theta^2-(\Omega+\ii\eta)^2/4}\right]}_{(a)}\nn\\
                                       &+\underbrace{\IM\left[\int_{-\delta}^{+\delta}\dd{\varphi} \int_0^{\Lambda_{\text{b}} }\dd{\ve}\frac{1}{\sqrt{\ve^2+\Delta_{\rm min}^2}}\frac{\Delta_{\rm min}^2}{\ve^2+\Delta_{\rm min}^2+16\Delta_4\Delta_{\rm min}\varphi^2-(\Omega+\ii\eta)^2/4}\right]}_{(b)}.
\end{align}
The term $(a)$ integrates in exactly the same manner as \cref{eq:aterm}, leading to
\begin{align}\label{eq:aterm_anis}
  (a)&=\Delta_{\rm max} \IM\left[\int_{-\delta}^{+\delta} \dd{\theta}\int_0^\lambda \dd{\e}\frac{1}{\e^2+\Delta_{\rm max}^2-16\Delta_{\rm max}\Delta_4\theta^2-(\Omega+\ii\eta)^2/4}\right]\nn\\
     &=\frac{\sqrt{\Delta_{\rm max}}}{8\sqrt{\Delta_4}} \IM\left\{\int\dd{x}\dd{y}\frac{1}{x^2-y^2-2\Delta^2_{\rm max}[\Omega/(2\Delta_{\rm max})-1]-\ii\eta}\right\}\nn\\
     &=\frac{\pi}{8}\sqrt{\frac{\Delta_{\rm max}}{\Delta_4}}\ln\left[\frac{\lambda}{|\Omega/(2\Delta_{\rm max})-1|}\right].
\end{align}
The term $(b)$ has a form similar to \cref{eq:aterm_anis} but this time the integration is not over the saddle-stationary point [$x^2-y^2$ form, \cref{eq:2D2}], but over a parabolic-stationary point [$x^2+y^2$ form, \cref{eq:2D1}]. This gives
\begin{align}\label{eq:bterm_anis}
  (b)&=\Delta_{\rm min} \IM\left[\int_{-\delta}^{+\delta} \dd{\varphi}\int_0^\lambda \dd{\e}\frac{1}{\e^2+\Delta_{\rm min}^2+16\Delta_{\rm min}\Delta_4\varphi^2-(\Omega+\ii\eta)^2/4}\right]\nn\\
     &=\frac{\sqrt{\Delta_{\rm min}}}{8\sqrt{\Delta_4}} \IM\left\{\int \dd{x}\dd{y}\frac{1}{x^2+y^2-2\Delta^2_{\rm min}[\Omega/(2\Delta_{\rm min})-1]-\ii\eta}\right\},\nn\\
     &=\frac{\pi^2}{8}\sqrt{\frac{\Delta_{\rm min}}{\Delta_4}}\Theta(\Omega-2\Delta_{\rm min}).
\end{align}
\ewt
If the pole is not at low energies which happens when $\Omega\gg\Delta_{\rm max}$, the integration is evaluated in the same manner as \cref{eq:largeW}, differing only in the angular averaging which takes the form $\int \dd{\theta}[\Delta_0+\Delta_4\cos(2\theta)]^2$. The nodeless case can then be summarized as:
\begin{align}\label{eq:nodeless_aog_result}
  &\IM[\chi^{s\text{-nodeless}}_{\Aog}(\Omega)]\nonumber\\
  & = \begin{cases}
    \displaystyle 0 &\text{for }\Omega\ll2\Delta_{\rm min},\\
    \displaystyle \frac{\pi}{4}\sqrt{\frac{\Delta_{\rm min}}{\Delta_4}}\Theta(\Omega-2\Delta_{\rm min}) &\text{for }\Omega\approx2\Delta_{\rm min},\\[2ex]
    \displaystyle \frac{1}{4}\sqrt{\frac{\Delta_{\rm max}}{\Delta_4}}\ln\left[\frac{\lambda}{|\Omega/(2\Delta_{\rm max})-1|}\right] &\text{for }\Omega\approx2\Delta_{\rm max},\\[2ex]
    \displaystyle \frac{\pi}{2}\frac{4\Delta_0^2+2\Delta_4^2}{\Omega^2} &\text{for }\Omega\gg2\Delta_{\rm max}.
  \end{cases}
\end{align}
These asymptotic results are plotted against the exact result in \cref{fig:A1g_results_b}, in which we again observe an excellent agreement.

Moving now to the nodal case, we first note that there are three sets of stationary points. Two of them are maxima points with $\Delta_{{\rm max}_1}=\Delta_{\theta=\pi/4}=|\Delta_0-\Delta_4|$ and $\Delta_{{\rm max}_2}=\Delta_{\theta=0,\pi/2}=\Delta_0+\Delta_4$. The third is a set of two minima points where $\Delta_\theta=0$ for $\theta=\pi/4\pm\theta_0$, with $\sin(\theta_0)=\Delta_0/\Delta_4$. See the inset of \cref{fig:A1g_results_c} for a schematic of the structure of the order parameter. The two different values of the maxima will induce the same $\ln$ feature but at different values of $\Omega$. The nodal regions (of which there are two) will each contribute in the same manner as in the $d$-wave case. The contribution in this region will only differ by the coefficient of $(\theta-\pi/4)^2$ term. This is nothing but the square of the slope of $\Delta_\theta$ at the nodal point. In the $d$-wave case this was $\propto\Delta^2_2$, which now changes to $4(\Delta_4^2-\Delta_0^2)$. Thus, we can write the response for the nodal case as
\begin{align}\label{eq:nodal_aog_result}
  & \IM[\chi^{s\text{-nodal}}_{\Aog}(\Omega)]\nonumber\\
  &=\begin{cases}
    \displaystyle 2\times\frac{\pi}{8}\frac{\Omega}{2\sqrt{\Delta_4^2-\Delta_0^2}} & \text{for }\Omega\sim0,\\[2ex]
    \displaystyle \frac{1}{4}\sqrt{\frac{\Delta_{{\rm max}_1}}{\Delta_4}}\ln\left[\frac{\lambda}{|\Omega/(2\Delta_{{\rm max}_1})-1|}\right] & \text{for }\Omega\approx2\Delta_{{\rm max}_1},\\[2ex]
    \displaystyle \frac{1}{4}\sqrt{\frac{\Delta_{{\rm max}_2}}{\Delta_4}}\ln\left[\frac{\lambda}{|\Omega/(2\Delta_{{\rm max}_2})-1|}\right] & \text{for }\Omega\approx2\Delta_{{\rm max}_2},\\[2ex]
    \displaystyle \frac{\pi}{2}\frac{4\Delta_0^2+2\Delta_4^2}{\Omega^2} & \text{for }\Omega\gg2\Delta_{{\rm max}_2}.
  \end{cases}
\end{align}
The factor of $2$ in the first line arises due to there being two nodal points in the angular integration. An explicit association of these non-analytic features with the order parameter structure for anisotropic superconductors was not stated before to the best of our knowledge~\footnote{We mention here that in Ref.~\cite{Devereaux1995} a step jump feature was numerically evaluated while considering a mixed symmetry $s+d$ state, but the reason for the feature was never explained.}. Our prescription allows us to identify them unambiguously. The asymptotic results are shown in \cref{fig:A1g_results_c}. We see that, in addition to reproducing the singular features, we are able to also indicate the region of the Fermi surface that is responsible for these features. These are colour coded in the insets.

\subsection{Selection properties of the probe-related form factor}\label{Subsec:Selection}
Apart from gaining information about which regions of the integration are responsible for the universal features, we have also learned about the origin of these features. A nodal point region of the Fermi surface contributes a linear-in-$\Omega$ term, a minimum provides a step jump, while a maximum provides a $\ln$ peak. However, these contributions can change if the integration regions leading to these features are weighted differently. To demonstrate this, let us return to \cref{eq:chi1} and explore the cases for $C=\{\Bog,\Btg\}$. Note that $\gamma_C$ does not affect the poles; thus, the same regions as in the $C=\Aog$ cases would contribute. However, due to the $\theta$-dependence of the form factor, these regions would be weighted differently and, hence, alter the nature of the features. Let us explore such changes below.

\subsubsection{Raman correlation function in the $\Bog$ channel of a square lattice}
Here, $\gamma_{\Bog}=\cos(2\theta)$. This form factor maximally picks up the contributions from $\theta\sim 0,\pi/2$, while the contributions from $\theta\sim\pi/4$ are suppressed with a weight $(\theta-\pi/4)^2$. This variation does not qualitatively affect the response of isotropic superconductors, since all regions of the Fermi surface contribute the same feature ($1/\sqrt{\Omega/(2\Delta_0)-1}$), but with different weights, resulting only in a change in the overall scale of the response. However, interesting scenarios can arise in anisotropic superconductors (both $d$- and $s$-wave).

\begin{figure*}
  \centering
  \phantomsubfloat{\label{fig:b1g_a}}\phantomsubfloat{\label{fig:b1g_b}}\phantomsubfloat{\label{fig:b1g_c}}\vspace{-2\baselineskip}
  \includegraphics[width=1.0\linewidth]{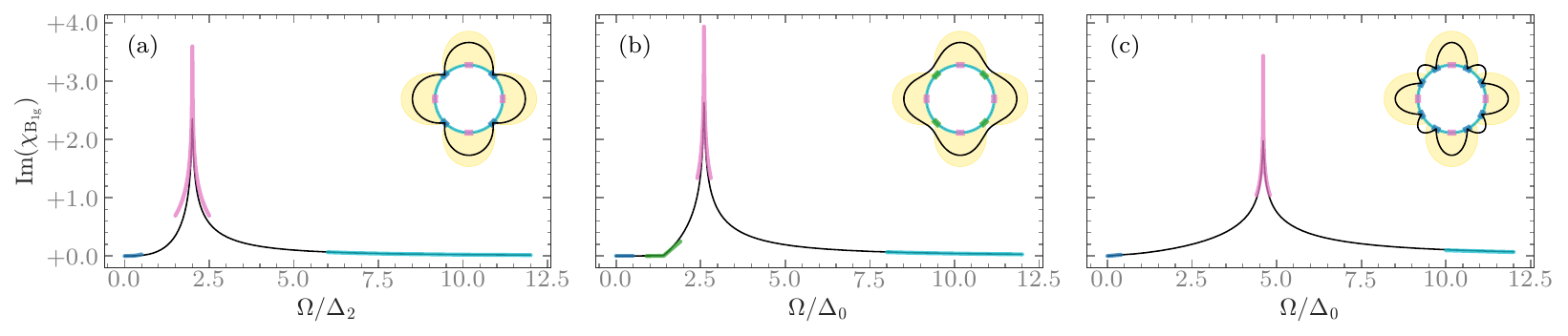}
  \caption{\textbf{Modification due to a $\Bog$ vertex form factor}: The asymptotic features of the bare Raman correlation function for $C=\Bog$ for (a) a $d$-wave, (b) a nodeless anisotropic \(s\)-wave and (c) a nodal anisotropic \(s\)-wave SC. The curves and insets follow the same scheme as \cref{fig:A1g_results}. When the maxima of \(\Delta_\theta\) and \(\gamma_{\Bog}\) coincide, the log peaks present in the isotropic vertex are picked up. However, if the form factor \(\gamma_{\Bog}\rightarrow0\), any feature is suppressed. When compared with \cref{fig:A1g_results}, we can clearly see the selection property of the vertex form factor.}
  \label{fig:b1g}
\end{figure*}

Let us first consider the $d$-wave case. We saw in Sec.~\ref{Subsec:dwave} that $\theta=0,\pi/2$ corresponded to the maximum of the pole and thus contributed a $\ln$ feature. This contribution will be picked up by the vertex $\gamma_{\Bog}$ with the weight factor $\cos^2(2\theta)\approx 1$ around these regions. Hence, it will not modify the $\Aog$ results around this feature. However, the $\theta\sim\pi/4$ region, which corresponded to the node and gave a response $\propto\Omega$, will now be weighted by $(\theta-\pi/4)^2 = \varphi^{2}$. Modifying the integration in \cref{eq:bterm} with this weight, we would get an extra factor of $\varphi^2$ in the numerator leading to the factor $r^2\cos^2(\xi)$ in the polar form:
\begin{align}\label{eq:bterm_B1g}
  (b) &= \frac{1}{4\Delta_0} \IM\Bigg[\int_0^\lambda r \dd{r}\int_0^{2\pi} \dd{\xi} \nonumber\\
      &\hphantom{= \frac{1}{4\Delta_0} \IM\Bigg[}\times\frac{1}{r}\frac{r^2\cos^2(\xi)}{r^2-(\Omega+\ii\eta)^2/4}\frac{r^2\cos^2(\xi)}{\Delta_0^2}\Bigg]\nn\\
      &=\frac{1}{4\Delta^3_0}\underbrace{\IM\left[\int_0^\lambda \dd{r}\frac{r^4}{r^2-(\Omega+\ii\eta)^2/4}\right]}_{(\Omega/2)^4\times\pi/\Omega~\text{from \cref{eq:1D2}}}\underbrace{\int_0^{2\pi} d\xi\cos^4(\xi)}_{3\pi/4}\nn\\
      &=\frac{3\pi^2}{32}\left(\frac{\Omega}{2\Delta_0}\right)^3.
\end{align}
That is, a linear behaviour is suppressed to a weaker cubic one. The two extra powers of $\Omega$ arise from the $\varphi^2$ contribution of the vertex $\gamma_{\Bog}$. Finally, the large frequency behaviour will also be affected, but only by the modification of the angular integration from $\int \dd{\theta}\cos^2(2\theta)\rightarrow\int \dd{\theta}\cos^4(2\theta)$. Collecting all these points, we can state the final result as
\begin{equation}\label{eq:d_bog_result}
  \IM[\chi^{d\text{-wave}}_{\Bog}(\Omega)]=\begin{cases}
    \displaystyle \frac{3\pi}{16}\left(\frac{\Omega}{2\Delta_2}\right)^3 &\text{for }\Omega\ll2\Delta_2,\\[2ex]
    \displaystyle \frac{1}{2}\ln\left[\frac{\lambda}{|\Omega/(2\Delta_2)-1|}\right] &\text{for }\Omega\approx2\Delta_2,\\[2ex]
    \displaystyle \frac{3\pi}{16}\frac{1}{[\Omega/(2\Delta_2)]^2} &\text{for }\Omega\gg2\Delta_2.
  \end{cases}
\end{equation}
These results are plotted against the numerical computation in \cref{fig:b1g_a}.

We can repeat the same analysis for anisotropic $s$-wave systems. For the nodeless case with \(\Delta_{4} < \Delta_{0}\), the stationary points are at $\theta\sim 0$ (maximum) and $\theta\sim \pi/4$ (minimum). The former case leads to $\gamma_{\Bog}\approx 1$ and thus does not change the result from the $\Aog$ counterpart. This leads to the expected $\ln$ peak when the pole $\rightarrow0$, i.e., $\Omega\rightarrow2(\Delta_0+\Delta_4)$. In the latter case $\gamma_{\Bog}\approx2(\theta-\pi/4)$, which modifies \cref{eq:bterm_anis} as:
\bwt
\begin{align}\label{eq:bterm_anisb1g}
  (b)&=\Delta_{\rm min}\IM\left[\int_{-\delta}^{+\delta} \dd{\theta}\int_0^\lambda \dd{\e}\frac{4\theta^2}{\e^2+\Delta_{\rm min}^2+16\Delta_{\rm min}\Delta_4\theta^2-(\Omega+\ii\eta)^2/4}\right]\nn\\
     &=\frac{1}{32\Delta_{\rm min}^{1/2}\Delta_4^{3/2}} \IM\left\{\int \dd{\varphi}\int \dd{\e}\frac{\varphi^2}{\e^2+\varphi^2-[(\Omega+\ii\eta)^2/4-\Delta_{\rm min}^2]}\right\}\nn\\
     &=\frac{\pi^2}{32}\left(\frac{\Delta_{\rm min}}{\Delta_4}\right)^{3/2}\left(\frac{\Omega}{2\Delta_{\rm min}}-1\right)\Theta(\Omega-2\Delta_{\rm min}).
\end{align}
\ewt
Observe that a step jump is suppressed to a linear onset from zero due to the vertex suppression. It should be noted that without the weight function of $\varphi^2$ in \cref{eq:bterm_anisb1g}, the result would have just been $\sim\Theta(\Omega-2\Delta_{\rm min})$. Because of $\varphi^2$, our power counting argument requires two powers of the pole $\sqrt{\Omega^2-4\Delta^2_{\rm min}}$ to be introduced. Indeed, the multiplicative factor of $\Omega^2-4\Delta^2_{\rm min}$ appears in \cref{eq:bterm_anisb1g}. But in the particular form above, the factor has been approximated by $4\Delta_{\rm min}(\Omega-2\Delta_{\rm min})$ since this is a local asymptotic form near $\Omega\sim 2\Delta_{\rm min}$. Finally, when the pole tends to some large value compared to the low-energy scale of the order parameter ($\Omega\gg2\Delta_{\rm max}$), we get a contribution similar to \cref{eq:largeW} but with the angle integration $\int \dd{\theta}\cos^2(2\theta)[\Delta_0+\Delta_4\cos(4\theta)]^2=\pi(\Delta_0^2+\Delta_4^2/2)$. Collecting these points, we get
\bwt
\begin{equation}\label{eq:nodeless_bog_result}
  \IM[\chi^{s\text{-nodeless}}_{\Bog}(\Omega)] = \begin{cases}
    \displaystyle 0 & \text{for }\Omega\ll2\Delta_{\rm min},\\
    \displaystyle \frac{\pi}{16}\left(\frac{\Delta_{\rm min}}{\Delta_4}\right)^{3/2}\left(\frac{\Omega}{2\Delta_{\rm min}}-1\right)\Theta(\Omega-2\Delta_{\rm min}) & \text{for }\Omega\approx2\Delta_{\rm min},\\[2ex]
    \displaystyle \frac{1}{4}\sqrt{\frac{\Delta_{\rm max}}{\Delta_4}}\ln\left[\frac{\lambda}{|\Omega/(2\Delta_{\rm max})-1|}\right] & \text{for }\Omega\approx2\Delta_{\rm max},\\[2ex]
    \displaystyle \frac{\pi}{8}\frac{4[\Delta_0^2+(\Delta_0+\Delta_4)^2]}{\Omega^2}& \text{for }\Omega\gg2\Delta_{\rm max}.
  \end{cases}
\end{equation}
\ewt
The comparison of these results with the exact ones are shown in \cref{fig:b1g_b}.

For the \emph{nodal} case with \(\Delta_{4} > \Delta_{0}\), there are three stationary points. Two of them are local maxima at $\theta=\pi/4$ ($\Delta_{\rm max_1}=\Delta_4-\Delta_0$) and at $\theta=0,\pi/2$ ($\Delta_{\rm max_2}=\Delta_4+\Delta_0$), and one minimum at $\theta=\pi/4\pm\arcsin(\Delta_0/\Delta_4)$. As before, the maximum at $\theta\sim0$ still yields the $\ln$ feature as $\gamma_{\Bog}$ does not affect this region. The second maximum at $\theta\sim\pi/4$ is now weighted by the small factor $(\theta-\pi/4)^2$. This completely washes out the expected $\ln$ singularity. Carrying out the integration by following the corollary in Appendix~\ref{Sec:App4}, one finds the result to be a non-universal number of $\mathcal{O}(1)$ and hence not a singular function of the external parameter. The minima points, which are at the nodes of the order parameter, are not weighted by any suppressing terms for the form factor of $\gamma_{\Bog}$ and hence preserve the linear onset behaviour. The large frequency behaviour leads to the same universal $1/\Omega^2$ behaviour weighted by the appropriate angular averaging. Collecting these results we get
\begin{align}\label{eq:nodal_bog_result}
  &\IM[\chi^{s\text{-nodal}}_{\Bog}(\Omega)] \nonumber\\
  &=\begin{cases}
    \displaystyle 2\times\frac{\pi}{16}\frac{\Omega}{2\Delta_4}\sqrt{\frac{\Delta_4-\Delta_0}{\Delta_4+\Delta_0}} & \text{for }\Omega\sim0,\\[2ex]
    \displaystyle \mathcal{O}(1) & \text{for }\Omega\approx2\Delta_{{\rm max}_1},\\[1ex]
    \displaystyle \frac{1}{4}\sqrt{\frac{\Delta_{{\rm max}_2}}{\Delta_4}}\ln\left[\frac{\lambda}{|\Omega/(2\Delta_{{\rm max}_2})-1|}\right] & \text{for }\Omega\approx2\Delta_{{\rm max}_2},\\[2ex]
    \displaystyle \frac{\pi}{8}\frac{4[\Delta_0^2+(\Delta_0+\Delta_4)^2]}{\Omega^2} & \text{for }\Omega\gg2\Delta_{{\rm max}_2}.
  \end{cases}
\end{align}
The comparison of these asymptotic features with the exact ones is shown in \cref{fig:b1g_c}.
\begin{figure*}
  \centering
  \phantomsubfloat{\label{fig:b2g_a}}\phantomsubfloat{\label{fig:b2g_b}}\phantomsubfloat{\label{fig:b2g_c}}\vspace{-2\baselineskip}
  \includegraphics[width=1.0\linewidth]{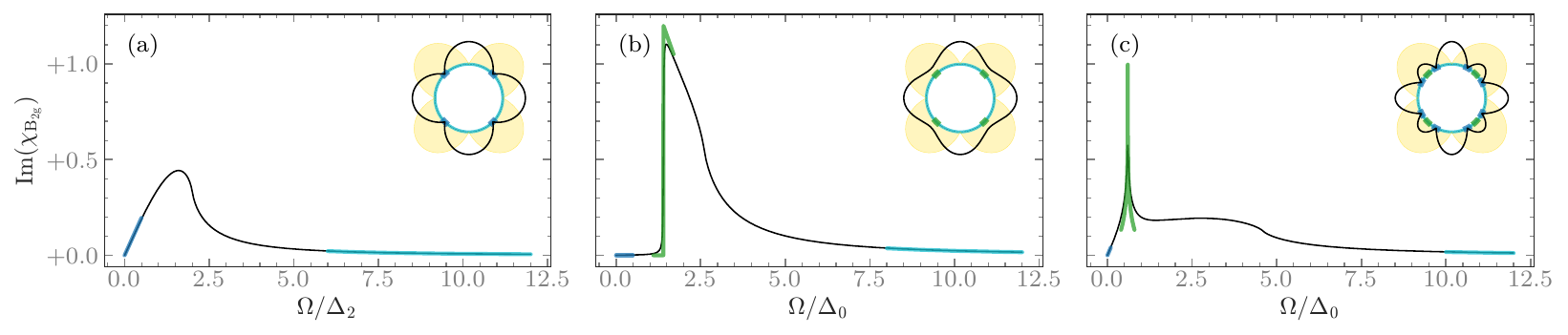}
  \caption{\textbf{Modification due to a $\Btg$ vertex form factor}: The bare Raman correlation function for $C=\Btg$ for (a) a $d$-wave, (b) a nodeless anisotropic \(s\)-wave and (c) a nodal anisotropic \(s\)-wave superconductor. 
    The curves and insets follow the same scheme as \cref{fig:A1g_results}. The lack of $\ln$ feature of $\Aog$ response in the \(d\)-wave case, (a), is because here \(\gamma_{\Btg}\rightarrow0\). However, similar to \Aog{} case, \cref{fig:A1g_results}, the jump is found in the nodeless anisotropic \(s\)-wave case, (b), as it is associated with regions where \(\gamma_{\Btg}\) remains finite. Finally, only one of the two $\ln$ peaks of $\Aog$ for the nodal case, (c), gets selected due to the $\Btg$ form factor.}
  \label{fig:b2g}
\end{figure*}

\subsubsection{Raman correlation function in the $\Btg$ channel of a square lattice}
For the irrep $C=\Btg$, we have $\gamma_{\Btg}=\sin(2\theta)$. This changes the relative alignment of the stationary points of the pole and the zeros of the form factor. We will not repeat the steps in this case and only state the results as the analysis is similar to the case with $C=\Bog$. Since the $\Btg$ form factor suppresses features arising from $\theta\sim 0,\pi/2$, we should expect some $\ln$ features that appeared in the $\Bog$ case to not appear here and some step jumps that did not show up in the $\Bog$ case to show up here. These differences are evident in the results below:
\begin{equation}\label{eq:d_btg_result}
  \IM[\chi^{d\text{-wave}}_{\Btg}(\Omega)]=\begin{cases}
    \displaystyle \frac{\pi}{4} \frac{\Omega}{2\Delta_2} &\text{for }\Omega\ll2\Delta_2, \\[2ex]
    \displaystyle \mathcal{O}(1) &\text{for } \Omega\approx2\Delta_2,\\[1ex]
    \displaystyle \frac{\pi}{16}{\frac{1}{[ \Omega/(2\Delta_0) ]^2}} &\text{for }\Omega\gg2\Delta_2;
  \end{cases}
\end{equation}
\begin{align}\label{eq:nodeless_btg_result}
  &\IM[\chi^{s\text{-nodeless}}_{\Btg}(\Omega)]\nonumber\\
  &=\begin{cases}
    \displaystyle 0 & \text{for } \Omega\ll2\Delta_{\rm min},\\
    \displaystyle \frac{\pi}{4}\sqrt{\frac{\Delta_{\rm min}}{\Delta_4}}\Theta(\Omega-2\Delta_{\rm min}) & \text{for } \Omega\approx2\Delta_{\rm min},\\[2ex]
    \displaystyle \mathcal{O}(1) & \text{for } \Omega\approx2\Delta_{\rm max},\\[1ex]
    \displaystyle \frac{\pi}{8}\frac{4[\Delta_0^2+(\Delta_0-\Delta_4)^2]}{\Omega^2}& \text{for } \Omega\gg2\Delta_{\rm max};
  \end{cases}
\end{align}
\bwt
\begin{equation}\label{eq:nodal_btg_result}
  \IM[\chi^{s\text{-nodal}}_{\Btg}(\Omega)]=\begin{cases}
    \displaystyle \frac{\pi}{8}\frac{\Omega}{2\Delta_4}\sqrt{\frac{\Delta_4+\Delta_0}{\Delta_4-\Delta_0}} & \text{for } \Omega \sim 0,\\[2ex]
    \displaystyle \frac{1}{4}\sqrt{\frac{\Delta_{{\rm max}_1}}{\Delta_4}}\ln\left[{\frac{\lambda}{|\Omega/(2\Delta_{{\rm max}_1})-1|}}\right] & \text{for } \Omega\approx2\Delta_{{\rm max}_1},\\[2ex]
    \displaystyle \mathcal{O}(1) & \text{for } \Omega\approx2\Delta_{\rm max_2},\\[1ex]
    \displaystyle \frac{\pi}{8}\frac{4[\Delta_0^2+(\Delta_0-\Delta_4)^2]}{\Omega^2} & \text{for } \Omega\gg2\Delta_{\rm max}.
  \end{cases}
\end{equation}
\ewt
These asymptotic results are plotted against the exact results in \cref{fig:b2g}. In line with the discussion above, we can verify in these figures that the features of $\Aog$ show up when $\gamma_C$ remains finite and do not when $\gamma_C\rightarrow0$. The profile of $\gamma_{\Btg}$ is shown in the insets of \cref{fig:b2g}.

Although the spectral line shapes in all these responses are quite distinct, we have hopefully convinced the reader that the different manifestations ultimately arise from universal behaviours of the response functions near the poles.

\bigskip

\paragraph*{Improving the asymptotic formulas:} There is one aspect in the case of nodeless \(s\) wave that is not reflected in the formulæ: the downward slope found at $\Omega\gtrsim2\Delta_{\rm min}$. To obtain this feature, we have to keep the next-order correction in $\theta$ to the $\gamma_{\Btg}$ form factor. Since this contribution comes from $\theta=\pi/4$, the correction is $\gamma_{\Bog}\approx 1-2(\theta-\pi/4)^2$. The integration becomes similar to \cref{eq:bterm_anis}, but with a weight factor of $1-4(\theta-\pi/4)^2$. Thus, we should expect a correction to the step function. Indeed, writing the expression close to the pole, we get:
\bwt
\begin{align}\label{eq:bterm_anisX}
  (b) &=\Delta_{\rm min} \IM\left[\int_{-\delta}^{+\delta} \dd{\varphi}\int_0^\lambda \dd{\e}\frac{1-4\varphi^2}{\e^2+\Delta_{\rm min}^2+16\Delta_{\rm min}\Delta_4\varphi^2-(\Omega+\ii\eta)^2/4}\right]\nn\\
      &=\frac{\sqrt{\Delta_{\rm min}}}{8\sqrt{\Delta_4}} \IM\left\{\int \dd{x}\dd{y}\frac{1-y^2/(4\Delta_{\rm min}\Delta_4)}{x^2+y^2-2\Delta^2_{\rm min}[\Omega/(2\Delta_{\rm min})-1]-\ii\eta},\right\}\nn\\
      &=\frac{\pi^2}{8}\sqrt{\frac{\Delta_{\rm min}}{\Delta_4}}\Theta(\Omega-2\Delta_{\rm min})\left[1-\frac{\Omega/2-\Delta_{\rm min}}{4\Delta_4}\right],
\end{align}
\ewt
where $\phi\equiv \theta-\pi/4$ as before. This is what is plotted in \cref{fig:b2g_b} for \(\Omega \approx 2\Delta_{\mathrm{min}}\). As is evident, the prescription allows one to choose the appropriate order of corrections to the asymptotic terms depending on the extent of improvement one seeks.

\section{Application to density of states}\label{Sec:Applications1}
Having learned about possible universal features and their origins for the Raman correlation function, it should be clear that our prescription is not limited to this. It can be readily extended to calculate other relevant quantities such as the density of states (DOS). The results for DOS in the systems we will present below are well known, but we use them as examples to demonstrate the validity as well as the ease of using our prescription. We start from the general definition~\cite{mahan_2011_CondensedMatterNutshell}:
\begin{equation}\label{eq:dosdef}
  g_{d\text{D}}(E)=V^d\int_{\bk}\delta(E-\ve_{\bk}),
\end{equation}
where $d$ is the spatial dimension, $\int_\bk\equiv\int \dd[d]{k}/(2\pi)^d$, $V^d$ is the $d$-dimensional volume, and $\varepsilon_\bk$ are the energy levels of the system. We can also rewrite this equation as
\begin{equation}\label{eq:dosdef2}
  g_{d\text{D}}(E)\stackrel{\eta\rightarrow0}{=}-\frac{V^d}\pi \IM\left(\int_{\bk}\frac1{E-\ve_{\bk}+\ii\eta}\right),
\end{equation}
which brings it to the familiar form that we have been using.

\subsection{Density of states of an electron gas in 1, 2 and 3 dimensions}

\begin{figure*}[ht]
  \centering
  \phantomsubfloat{\label{fig:dos_isotropic_a}}\phantomsubfloat{\label{fig:dos_isotropic_b}}\phantomsubfloat{\label{fig:dos_isotropic_c}}\vspace{-2\baselineskip}
  \includegraphics[width=1.0\linewidth]{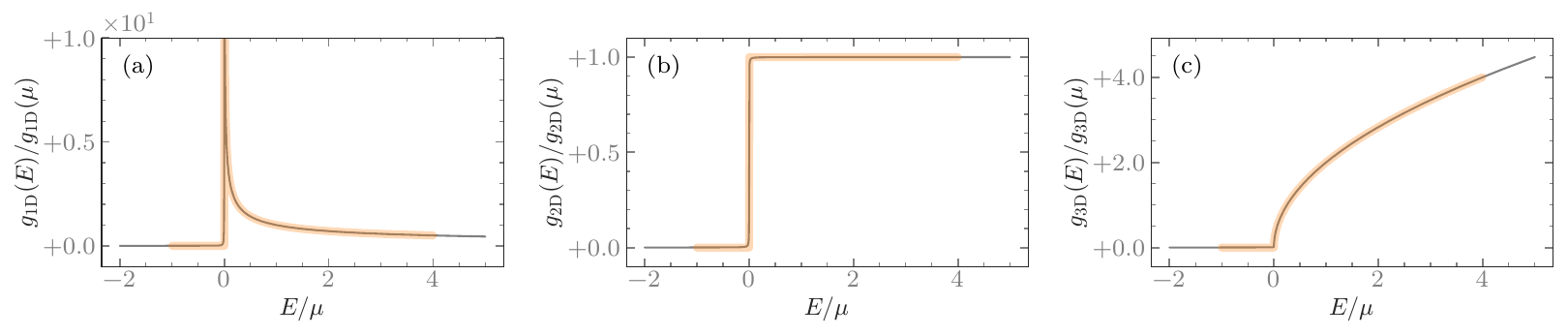}
  \caption{\textbf{Asymptotic analysis of DOS of free-electron systems:} The asymptotic results (light-coloured thick lines) for the integration against exact calculations (black thin line) for the density of states for (a) 1D, (b) 2D and (c) 3D isotropic systems, computed using the definition, \cref{eq:dosdef2}, and  normalized by its value at Fermi level \(\mu\). The asymptotic results accurately reproduce the features in the DOS over the whole range.}
  \label{fig:dos_isotropic}
\end{figure*}
For an electron gas, the energy dispersion is isotropic, $\ve=\hbar^2\bk^2/(2m)$. Certainly, one can exactly integrate over the $\delta$ function in \cref{eq:dosdef} to get the DOS, but this integration is not always possible (e.g., if the system is a lattice). However, should our prescription work, we should be able to deduce the result simply by considering the stationary points and locally integrating around them. In 1D for a free-electron system, the stationary point is at $k=0$, and the integral can be cast in the following manner:
\begin{align}\label{eq:1D}
  g_{\text{1D}}(E)&=\frac{L}\pi \IM\left[\int_{-\infty}^{+\infty}\frac{\dd{k}}{2\pi}\frac1{\hbar^2k^2/(2m)-E-\ii\eta}\right]\nn\\
                  &=\frac{\sqrt{2m}L}{2\pi^2\hbar} \underbrace{\IM\left(\int_{-\infty}^{+\infty} \dd{k}\frac1{k^2-E-\ii\eta}\right)}_{\mathclap{=2\times\pi\Theta(E)/(2\sqrt{E})~\text{from \cref{eq:1D2}}}}\nn\\
                  &=\frac{\sqrt{m}L}{\sqrt{2}\pi\hbar}\frac{1}{\sqrt{E}}\Theta(E).
\end{align}
The factor of $2$ in the integration step arises from the fact that the integrand has two poles at $k=\pm\sqrt{E}$. Similarly, for the 2D free electron system (with the stationary point at the origin) we get:
\begin{align}\label{eq:2D}
  g_{\text{2D}}(E)&=\frac{A}\pi \IM\left[\int_{0}^\infty \frac{k\dd{k}}{2\pi}\int_0^{2\pi}\frac{\dd{\theta}}{2\pi}\frac1{\hbar^2k^2/(2m)-E-\ii\eta}\right]\nn\\
                  &=\frac{mA}{\pi^2\hbar^2} \underbrace{\IM\left(\int_{0}^\infty \dd{k}\frac{k}{k^2-E-\ii\eta}\right)}_{\mathclap{=\sqrt{E}\times\pi\Theta(E)/(2\sqrt{E})~\text{from \cref{eq:1D2}}}}\nn\\
                  &=\frac{mA}{2\pi\hbar^2}\Theta(E).
\end{align}
Here, a factor of $\sqrt{E}$ appears in the integral before simplification due to the extra $k$ in the numerator. This, however, does not affect the local integration around the pole, which is at $k=\sqrt{E}$, and just appears as a prefactor. Next, in 3D we get
\begin{align}\label{eq:3D}
  g_{\text{3D}}(E)&=\frac{V}\pi \IM\Bigg[\int_{0}^\infty \frac{k\dd{k}}{2\pi}\int_0^{2\pi}\frac {\dd{\phi}}{2\pi}\int_0^\pi\frac{\sin\theta \dd{\theta}}{2\pi} \nonumber\\
                  &\hphantom{=\frac{V}\pi \IM\Bigg[}\times\frac1{\hbar^2k^2/(2m)-E-\ii\eta}\Bigg]\nn\\
                  &=\frac{(2m)^{3/2}V}{2\pi^3\hbar^3} \underbrace{\IM\left(\int_{0}^\infty \dd{k}\frac{k^2}{k^2-E-\ii\eta}\right)}_{\mathclap{=(\sqrt{E})^2\times\pi\Theta(E)/(2\sqrt{E})~\text{from \cref{eq:1D2}}}}\nn\\
                  &=\frac{m^{3/2}V}{\sqrt{2}\pi^2\hbar^3}\sqrt{E}\Theta(E).
\end{align}
Here, a factor of $(\sqrt{E})^2$ appears in the non-simplified integral because of the extra $k^2$ factor in the numerator.

In this simple example, the asymptotic forms are, in fact, already the exact results. See \cref{fig:dos_isotropic} for the comparison. Certainly, this will not always be the case, but it demonstrates that the local asymptotic analysis can also give exact results.

\begin{figure*}
  \centering
  \phantomsubfloat{\label{fig:dos_lattice_a}}\phantomsubfloat{\label{fig:dos_lattice_b}}\vspace{-2\baselineskip}
  \includegraphics[width=0.8\linewidth]{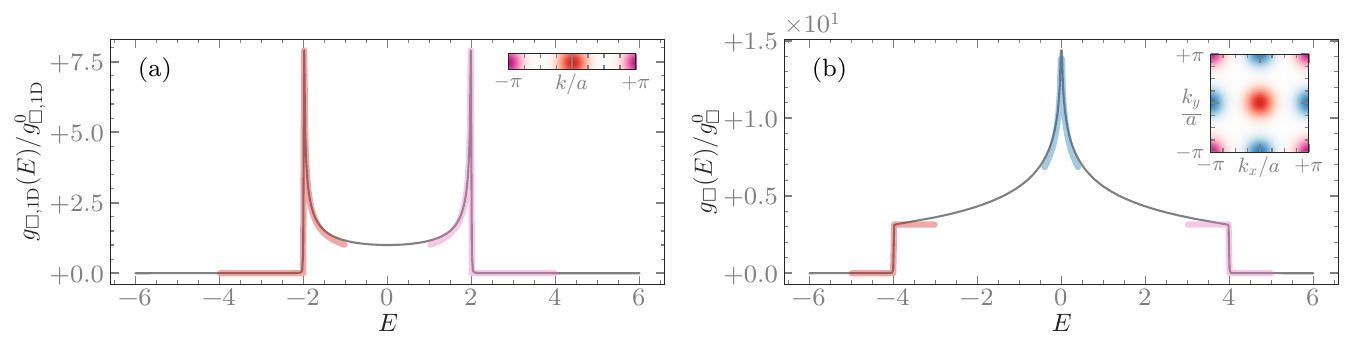}
  \caption{\textbf{Asymptotic analysis of DOS in lattice systems:} The asymptotic results for the DOS of a square lattice in (a) 1D and (b) 2D against exact results (black thin line). The asymptotic curves have been colour coded to match the regions of the Brillouin zone (insets). In 1D, (a), the square-root features arise from the centre ($k=0$) and edges ($k=\pm\pi$) of the Brillouin zone. In 2D, (b), the left (right) jump features arise from the \group{\Gamma} (\group{M}) high-symmetry points, while the $\ln$ peak from the \group{X} and \group{Y} points.}
  \label{fig:dos_lattice}
\end{figure*}

\subsection{van Hove singularities in band structures}
Let us now consider a more challenging scenario of the energy dispersion of a square lattice with nearest-neighbour hopping, $\ve_{\bk}=-2t[\cos(ak_x)+\cos(ak_y)]$, where \(a\) is the lattice parameter. The starting expression for DOS is
\begin{align}\label{eq:DOSsq}
  g_{\squarelattice}(E)=-\frac{A}{\pi} \IM\left\{\int_{\bk}\frac{1}{E+2t[\cos(ak_x)+\cos(ak_y)]+\ii\eta}\right\}.
\end{align}
The stationary points (obtained from $\partial_{k_x}\ve_{\bk}=0$ and $\partial_{k_y}\ve_{\bk}=0$) are at the $\group{\Gamma}$ [$(0,0)$], $\group{M}$ [$(\pm\pi,\pm\pi)$], $\group{X}$ [$(\pm\pi,0)$] and $\group{Y}$ [$(0,\pm\pi)$] points of the Brillouin zone. The $\group{\Gamma}$- and $\group{M}$-points are parabolic-stationary points and hence should result in a step jump (Sec.~\ref{Sec:Applications2}). These stationary points occur at different energies and thus we should expect the step jumps at each of these energies. The explicit steps of the prescription would work out in the following way:
\paragraph*{Feature from the $\group{\Gamma}$-point at $E=-2t$.} The integration takes the form
\begin{align}\label{eq:DOSsq2}
  g_{\squarelattice}(E\approx-2t)&=-\frac{A}{2\pi^2}\underbrace{\IM\left(\int_0^\delta k \dd{k}\frac{1}{E+2t-a^2tk^2+\ii\eta}\right)}_{=-\Theta(E+2t)\pi/(2a^2t)}\nn\\
                                 &=\frac{A}{4a^2\pi t}\Theta(E+2t).
\end{align}
\paragraph*{Feature from the $\group{M}$-points at $E=2t$.} Since the four $\group{M}$-points are equivalent, we can combine them into one. This leads to:
\begin{align}\label{eq:DOSsq3}
  g_{\squarelattice}(E\approx2t)&=-\frac{A}{2\pi^2}\underbrace{\IM\left(\int k \dd{k}\frac{1}{E-2t+a^2tk^2+\ii\eta}\right)}_{=-\Theta(2t-E)\pi/(2a^2t)}\nn\\
                                &=\frac{A}{4a^2\pi t}\Theta(2t-E).
\end{align}
\paragraph*{Feature from $\group{X}/\group{Y}$-points at $E=0$:} The $\group{X}/\group{Y}$ points are saddle-stationary points and hence should produce a $\ln$ feature. There are two of each points which can be combined into one $\group{X}$ and one $\group{Y}$ point, each producing such a feature. Explicitly:
\begin{align}\label{eq:DOSsq4}
  & g_{\squarelattice}(E\approx0)\nonumber\\
  &=-\frac{A}{4\pi^3}\Bigg\{\underbrace{\IM\left[\int \dd{k_x}\dd{k_y}\frac{1}{E-a^2t(k_x^2-k_y^2)+\ii\eta}\right]}_{=-\pi\ln(\lambda/|E|)/(a^2t)}\nonumber\\
  & \hphantom{=-\frac{A}{4\pi^3}\Bigg\{} + \underbrace{\IM\left[\int \dd{k_x}\dd{k_y}\frac{1}{E+a^2t(k_x^2-k_y^2)+\ii\eta}\right]}_{=-\pi\ln(\lambda/|E|)/(a^2t)}\Bigg\}\nn\\
  &=\frac{A}{2\pi^2a^2t}\ln\left(\frac{\lambda}{|E|}\right).
\end{align}

These results are plotted against an exact numerical calculation in \cref{fig:dos_lattice}.

\section{Conclusion}\label{Conclusion}
In this work, we have used a stationary-point analysis to come up with a prescription of rules to extract spectral features of response functions. We applied it to compute certain correlation functions that are common to the calculation of the Raman response of superconductors, and also to calculations of density of states in lattice systems. In both cases, we demonstrated how effectively the prescription reproduces rather involved features of a spectral function without an explicit computation. In doing so, we are also able to identify the location of these features as well as the regions of the integration space that lead to it.

We used this prescription to point out characteristic features of Raman correlation functions that arise due to the anisotropy of the superconducting order parameter. In particular, we showed on general grounds that the nodal regions of the order parameter yield a response linear in frequency, the minima regions lead to a step jump at the energies corresponding to the order-parameter minima, and the maxima regions lead to a $\ln$ singularity at the energies corresponding to the order-parameter maxima. These intrinsic features can get additionally modified by the particular choice of the probe. We showed that for every zero associated with the probe-related vertex, one gets an additional power of the pole-related singularity. This modifies the original singularity leading to a different manifestation of it. These singularities have been shown to originate from the same universal nonanalytic behaviour that is dictated by the integration around parabolic- and saddle-like stationary points. We argued that the different manifestations effectively serve as selection rules for a particular nonanalyticity to appear in a given response (compare \cref{fig:A1g_results} with \cref{fig:b1g} and \cref{fig:b2g}).

There can be many uses of this prescription. It provides us with some guiding principle to interpret spectral data in the spectroscopy of 2D materials, like Raman scattering, resonant inelastic X-ray scattering, or even the THz pump-probe experiments, where such correlation functions are ubiquitous. The technique can be used to identify the nature of singularities for algorithmic improvements in certain numerical methods \cite{benek-lins_maiti:2024:numerics}. Finally, noting the rise in the use of symbolic regression in physics-inspired learning algorithms~\cite{Keren2023}, one can imagine that building an association of resulting singular features with particular structures of the integrand can be extremely beneficial for the learning/training process. As a closing remark we emphasize that many-body correlations can further introduce its own set of nonanalytic features~\cite{Chubukov1993,Chubukov2003} that we do not address here, but anticipate that it should prove useful even in those scenarios as they also involve integration over poles.

\paragraph*{Acknowledgments:}
This work was funded (S.M.) by the Natural Sciences and Engineering Research Council of Canada (NSERC) Grant No. RGPIN-2019-05486 and partially (I.B.-L.) via the NSERC's QSciTech CREATE programme. Also, this research was enabled in part by the computing resources provided by Calcul Québec and the Digital Research Alliance of Canada. J.D.\ was in Concordia University when the initial results of the work were obtained. Lastly, the authors would like to extend their gratitude to Dmitrii Maslov who instiled in them the belief that if there is something interesting, one hardly ever needs a brute force computation to discover it.

\appendix
\section{1D integration of $1/(x-z)$}\label{Sec:App1}
Let us consider the integration
\begin{equation}\label{eq:b1}
  I(z)=\int_a^b \dd{x} \frac{1}{x-z-\ii\eta},
\end{equation}
where $\eta\rightarrow0$ and $z \in \mathbb{R}$. This is readily doable, yielding
\begin{equation}\label{eq:b2}
  I(z)=\ln\left(\frac{b-z-\ii\eta}{a-z-\ii\eta}\right).
\end{equation}
Observe that $\IM[I(z)]$ is $\pi$ if $a<z<b$ and zero otherwise. The fact that it is independent of the limits suggests that the result can be obtained from local integration around the pole. Indeed, by setting $a=z-\delta$ and $b=z+\delta$ (with $\delta>0$), we get
$$I(z)=\ln\left(\frac{+\delta-\ii\eta}{-\delta-\ii\eta}\right).$$
Here we have two small parameters: $\delta$, as it marks a small region around the poles, and $\eta$, which is small by definition. Since both are arbitrary choices, the inherent question here is about which parameter is smaller. This is resolved from the following perspective. The parameter $\eta$, though arbitrary, is fixed once the integration is identified. The parameter $\delta$, on the other hand, is under our control. The approach here would be that we can keep it as narrow or as wide we need, in relation to $\eta$, to pick up a universal contribution. This choice is available because the original integration is not limited to an integration around the pole, but covers a much larger range. It is clear that if $\delta\ll\eta$ then $I(z)\rightarrow0$. This means that we need to cover a larger integration region to pick up a contribution. Indeed, in the limit $\delta\gg\eta$ we get $I(z)=i\pi$. Thus, we are led to
\begin{equation}\label{eq:pole1}
  \IM[I(z)]=\IM\left(\int_{z-\delta}^{z+\delta}\dd{x}\frac{1}{x-z-\ii\eta}\right)=\pi \sgn(\eta),
\end{equation}
with the entire result coming just from the pole.

\section{1D integration of $1/(x^2-z)$}\label{Sec:App2}
Let us consider the integration
\begin{equation}\label{eq:a1}
  I(z)=\int_a^b \dd{x}\frac{1}{x^2-z-\ii\eta},
\end{equation}
where $\eta\rightarrow0$ and $z \in \mathbb{R}$. The integrand can be written as
$$\frac1{x^2-z}=\frac1{2(\sqrt{z}+\ii\tilde \eta)}\left(\frac{1}{x-\sqrt{z}-\ii\tilde\eta}-\frac1{x+\sqrt{z}+\ii\tilde\eta}\right),$$ where $\tilde\eta=\eta/(2\sqrt{z})$. This leads to a straightforward integration leading to
\bwt
\begin{equation}\label{eq:a2}
  I(z)=\begin{cases}
    \displaystyle \frac1{2(\sqrt{z}+\ii\tilde\eta)}\left[\ln\left(\frac{b-\sqrt{z}-\ii\tilde\eta}{b+\sqrt{z}+\ii\tilde\eta}\right)-\ln\left(\frac{a-\sqrt{z}-\ii\tilde\eta}{a+\sqrt{z}+\ii\tilde\eta}\right)\right] & \text{if $z>0$}\\[2ex]
    \displaystyle \frac1{\sqrt{-z}}\left[\arctan\left(\frac{b}{\sqrt{-z}}\right)-\arctan\left(\frac{a}{\sqrt{-z}}\right)\right] & \text{if $z<0$}.
  \end{cases}
\end{equation}
\ewt
In fact, if we extend the functions to the complex plane, both lines serve as interchangeable results and mean the same thing. Along the real axis, observe that if $z<0$, there is no imaginary part. Also, if $z>0$ and $a<\sqrt{z}<b$, then the imaginary part is just $\pi\sgn(\eta)$, independent of the limits. This is again suggestive of the fact that the universal result will be independent of the limits of integration. Thus, we can integrate locally around the poles and obtain
\begin{align}\label{eq:a3}
  \IM[I(z)]&=\IM\left[\int_{\sqrt{z}-\delta}^{\sqrt{z}+\delta}\dd{x}\frac{1}{x^2-z-\ii\eta}\Theta(z)\right]\nn\\
           &=\frac1{2\sqrt{z}}\ln\left(\frac{+\delta-\ii\tilde\eta}{-\delta-\ii\tilde\eta}\right)\Theta(z)\nn\\
           &= \frac{\pi}{2\sqrt{z}}\Theta(z) \sgn(\eta).
\end{align}
Here, like before, it is understood that we need an integration region $\delta\gg\eta$.

\section{2D integration of $1/(x^2+y^2-z)$}\label{Sec:App3}
From here on, we will only state the result for the imaginary part of integration around the poles. Let us consider the integration
\begin{equation}\label{eq:a4}
  I(z)=\int \dd{x}\dd{y}\frac{1}{x^2+y^2-z-\ii\eta},
\end{equation}
where $\eta\rightarrow0$, $z\in\mathcal{R}$ and the integration region is assumed to contain the pole. If there is a universal part to $\IM[I(z)]$, then it would be independent of the limits of integration. We can thus, conveniently, change the integration over polar coordinates \((r, \theta)\) without worrying about the limits (assuming the pole to be contained within the limits, of course) to write
\begin{align}\label{eq:a5}
  \IM[I(z)]&=\IM\left(\int_0^{2\pi} \dd{\theta}\int_{\sqrt{z}-\delta}^{\sqrt{z}+\delta} r \dd{r}\frac{1}{r^2-z-\ii\eta}\right)\nn\\
           &=\int_0^{2\pi} \dd{\theta} \sqrt{z}\underbrace{\IM\left(\int_{\sqrt{z}-\delta}^{\sqrt{z}+\delta}\dd{r}\frac{1}{r^2-z-\ii\eta}\right)}_{=(\pi/2\sqrt{z})\Theta(z)\sgn(\eta)~\text{from \cref{eq:a3}}}\nn\\
           &=\frac\pi2\Theta(z) \sgn(\eta)\int_0^{2\pi}\dd{\theta}\nn\\
           &=\pi^2 \Theta(z) \sgn(\eta).
\end{align}
In the second line, we pulled out the $r$ and replaced it with the value at the pole. In the penultimate line, we have retained the angle integration to emphasize that the numerator could contain additional functions of $\theta$ that will simply need to be integrated over.

\paragraph*{Corollary.} It is now easy to see that an integration of the type
\begin{equation}\label{eq:a4coll}
  I(z)=\int \dd{x}\dd{y} \frac{x^my^n}{x^2+y^2-z-\ii\eta}
\end{equation}
is also easily done if we are interested only in the imaginary part. The pole contribution remains the same, leading to
\begin{align}\label{eq:a4colla}
  &\IM[I(z)]\nonumber\\
  &=\IM\left(\int \dd{x}\dd{y}\frac{x^my^n}{x^2+y^2-z-\ii\eta}\right)\nn\\
  &= \frac{\pi}{2} (\sqrt z)^{m+n}\Theta(z) \sgn(\eta)\int_0^{2\pi}\dd{\theta} \cos^m(\theta)\sin^n(\theta)
\end{align}

\section{2D integration of $1/(x^2-y^2-z)$}\label{Sec:App4}
Let us consider the integration
\begin{equation}\label{eq:a6}
  I(z)=\int \dd{x}\dd{y}\frac{1}{x^2-y^2-z-\ii\eta},
\end{equation}
where $\eta\rightarrow0$ and $z\in\mathcal{R}$. As before, it is convenient to use polar coordinates, which leads us to
\begin{align}\label{eq:a7}
  &\IM[I(z)] \nonumber\\
  &=4 \IM\left[\int_0^{\pi/2} \dd{\theta}\int_0^\Lambda r \dd{r}\frac{1}{r^2\cos(2\theta)-z-\ii\eta}\right]\nn\\
  &=4 \IM\Bigg\{\Bigg[\underbrace{\int_0^{\pi/4-\delta}}_{(a)}+\underbrace{\int_{\pi/4-\delta}^{\pi/4+\delta}}_{(b)}+\underbrace{\int^{\pi/2}_{\pi/4+\delta}}_{(c)}\Bigg]\dd{\theta}\int_0^\Lambda r \dd{r}\nonumber\\
  &\hphantom{=4 \IM\Bigg\{}\times\frac{1}{r^2\cos(2\theta)-z-\ii\eta}\Bigg\}.
\end{align}
Here, $\Lambda$ is some upper limit of the $r$-integration that will not be relevant near the singularity. It is evident that if $z>0$, the pole is picked up by the integral $(a)$ and the lower half of the integral $(b)$. Also, if $z<0$, the pole is picked up by the upper half of the integral $(b)$ and the integral $(c)$. Thus, unlike the previous case, $z$ contributes to the result independently of its sign. Starting with $z>0$, the relevant contribution is
\bwt
\begin{align}\label{eq:a8}
  &\int_0^{\pi/4-\delta}\frac{\dd{\theta}}{\cos(2\theta)}\underbrace{\IM\left[\int_0^\Lambda \dd{r}\frac{r}{r^2-(z+\ii\eta)/\cos(2\theta)}\right]}_{=(\pi/2) \sgn(\eta)~\text{from \cref{eq:a5}}}+\IM\left[\int^{\pi/4}_{\pi/4-\delta}\dd{\theta}\int_0^\Lambda \dd{r}\frac{r}{r^2\cos(2\theta)-z-\ii\eta}\right]\nn\\
  &=\frac{\pi}{2}\int_0^{\pi/4-\delta}\frac{\dd{\theta}}{\cos(2\theta)}+\IM\left(\int_0^\delta \dd{\theta}\int_0^\Lambda \dd{r}\frac{r}{2r^2\theta-z-\ii\eta}\right).
\end{align}
\ewt
Once again, $\delta$ has to be chosen appropriately large, compared to \(\eta\), to pick up the finite contribution of the pole. The radial integration in the second term is the same as \cref{eq:a5} if $\theta>z/(2\Lambda^2)$, otherwise it is zero. This is just the condition to ensure that pole is found within the integration limits. This means that the angular integration in the second term has a lower limit at $z/(2\Lambda^2)$.  This leads to \cref{eq:a8} becoming
\begin{equation}\label{eq:a9}
  \frac{\pi}{2}\int_0^{\pi/4-\delta}\frac{\dd{\theta}}{\cos(2\theta)}+\frac{\pi}2\int_{z/(2\Lambda^2)}^\delta \frac{\dd{\theta}}{2\theta} \sgn(\eta).
\end{equation}
It is easy to check that the two terms produce mutually cancelling terms at the upper limits. The lower limit of the first term is not diverging and produces a result of $\mathcal{O}(1)$, but the lower limit of the second term produces $\pi\ln(\lambda/z)/4$, where $\lambda=\sqrt 2\Lambda$ is an undetermined constant. In the limit $z \rightarrow 0$, this produces a divergent contribution $\sim -\pi\ln z/4$. While we get the correct functional form, due to its singular nature, we cannot determine the $\mathcal{O}(1)$ constant using this method. However, these constants are formally negligible against the infinite singular contribution.

Following similar steps, it is easy to show that when $z<0$, the pole contribution is picked up by the terms $(b)$ and $(c)$ in exactly the same manner as above with the following changes in the signs in some of the terms:
\bwt
\begin{align}\label{eq:a10}
  &\int_{\pi/4+\delta}^{\pi/2}\frac{\dd{\theta}}{\cos(2\theta)}\IM\left[\int_0^\Lambda \dd{r}\frac{r}{r^2-(z+\ii\eta)/\cos(2\theta)}\right]+ \IM\left[ \int_{\pi/4}^{\pi/4+\delta}\dd{\theta}\int_0^\Lambda \dd{r}\frac{r}{r^2\cos(2\theta)-z-\ii\eta}\right]\nn\\
  &\stackrel{\theta\rightarrow\pi/2-\theta}{=}-\int^{\pi/4-\delta}_{0}\frac{\dd{\theta}}{\cos(2\theta)}\underbrace{ \IM\left[\int_0^\Lambda \dd{r}\frac{r}{r^2+(z+\ii\eta)/\cos(2\theta)}\right]}_{=(\pi/2) \sgn(-\eta)~\text{from \cref{eq:a5}}}-\IM\left[\int^{\pi/4}_{\pi/4-\delta}\dd{\theta}\int_0^\Lambda \dd{r}\frac{r}{r^2\cos(2\theta)+z+\ii\eta}\right]\nn\\
  &\stackrel{\phantom{\theta\rightarrow\pi/2-\theta}}{=}\frac{\pi}{2}\int_0^{\pi/4-\delta}\frac{\dd{\theta}}{\cos(2\theta)}- \IM\left[\int_0^\delta \dd{\theta}\int_0^\Lambda \dd{r}\frac{r}{2r^2\theta+z+\ii\eta}\right]\nn\\
  &\stackrel{\phantom{\theta\rightarrow\pi/2-\theta}}{=}\frac{\pi}{2}\int_0^{\pi/4-\delta}\frac{\dd{\theta}}{\cos(2\theta)}+\frac{\pi}{2}\int_{-z/(2\Lambda^2)}^\delta \frac{\dd{\theta}}{2\theta} \sgn(\eta).
\end{align}
\ewt
The last line is the same result as \cref{eq:a9}, since \(z<0\). Thus, by combining all the results, we can write
\begin{equation}\label{eq:a11}
  \IM[I(z)]\approx \pi\ln\left(\frac{\lambda}{|z|}\right) \sgn(\eta).
\end{equation}

\paragraph*{Corollary.} It is now easy to see that an integration of the type
\begin{equation}\label{eq:a4coll2}
  I(z)=\int \dd{x}\dd{y} \frac{x^my^n}{x^2-y^2-z-\ii\eta}
\end{equation}
is also easily done if we are interested only in the imaginary part. In polar coordinates \((r, \xi)\), the integration now looks like
\begin{align}\label{eq:a4collab}
  \IM[I(z)]&=\IM\left[\int \dd{\xi} \dd{r}\frac{r^{m+n+1}\cos^m(\xi)\sin^n(\xi)}{r^2\cos(2\xi)-z-\ii\eta}\right]\nn\\
           &=\int_{\pi/4+z} \dd{\xi}\frac{\pi z^{(m+n)/2}\cos^m(\xi)\sin^n(\xi)}{2[\cos(2\xi)]^{^{(m+n)/2+1}}}\nn\\
           &=\frac{\pi z^{(m+n)/2}}{4 2^{m+n}}\int_{z} \dd{\varphi}\frac{1}{\varphi^{(m+n)/2+1}}\nn\\
           &=\frac{\pi}{2(m+n)2^{m+n}}+\ldots\,.
\end{align}
The ``\(\ldots\)'' represents terms sensitive to the upper limit of the integration, which therefore are non-universal. This is just some background non-zero contribution that does not introduce any singular feature from the pole. While integration over a saddle point produced a log singularity, having any other power of the integration variable in the numerator completely removes it.

\section{Integration around non-stationary points}\label{Sec:App5}
Let us consider the integration
\begin{equation}\label{eq:a4coll3}
  I(z)=\int \dd{x}\dd{y} \frac{1}{x^2+y-z-\ii\eta},
\end{equation}
where $\eta\rightarrow0$, $z\in\mathcal{R}$ and one of the variables is not at the stationary point. Performing the $y$ integration in the range between $\pm\Lambda$ we get
\begin{equation}\label{eq:a4coll4}
  I(z)=\int \dd{x} \ln\left(\frac{x^2+\Lambda-z-\ii\eta}{x^2-\Lambda-z-\ii\eta}\right).
\end{equation}
The imaginary part of the above expression will only survive if $x^2-z+\Lambda$ and $x^2-z-\Lambda$ have opposite signs. This happens if $x\in(\sqrt{z-\Lambda},\sqrt{z+\Lambda})$. The imaginary part itself is then simply $\pi$, leading to the result $\pi(\sqrt{z+\Lambda}-\sqrt{z-\Lambda})$. This goes to zero if $\Lambda\rightarrow0$. That is, we do not pick up any universal contribution from the integration around the pole over a region that does not have a stationary point.

\end{document}